\documentstyle[epsfig,aps,prb,twocolumn,floats,eqsecnum]{revtex}

\begin{document}

\newcommand{\gsim}{
\,\raisebox{0.35ex}{$>$}
\hspace{-1.7ex}\raisebox{-0.65ex}{$\sim$}\,
}

\newcommand{\lsim}{
\,\raisebox{0.35ex}{$<$}
\hspace{-1.7ex}\raisebox{-0.65ex}{$\sim$}\,
}

\newcommand{\const}{ {\rm const} }
\newcommand{\arctanh}{ {\rm arctanh} }

\newcommand{\threehalf}{\mbox{\scriptsize 
\raisebox{1.5mm}{3}\hspace{-2.7mm}
\raisebox{0.5mm}{$-$}\hspace{-2.8mm}
\raisebox{-0.9mm}{2}\hspace{-0.7mm}
\normalsize }}

\bibliographystyle{prsty}

\title{ \begin{flushleft}
{\small 
PHYSICAL REVIEW B 
\hfill $\qquad$
VOLUME 
{\normalsize 59}, 
NUMBER
{\normalsize 1}
\hfill 
{\normalsize 1}
JANUARY
{\normalsize 1999}-I, 
{\normalsize 443-456}
}\\
\end{flushleft}  
Classical spin liquid: Exact solution for the infinite-component
antiferromagnetic model on the $\bbox{kagom\mathaccent19 e}$ lattice
}

\author{
D.~A. Garanin \cite{e-gar} and Benjamin~Canals \cite{e-can}
}

\address{
Max-Planck-Institut f\"ur Physik komplexer Systeme, N\"othnitzer Strasse 38,
D-01187 Dresden, Germany\\
\smallskip
{\rm (Received 22 May 1998) }
\bigskip\\
\parbox{14.2cm}
{\rm
Thermodynamic quantities and correlation functions (CF's) of the classical
antiferromagnet on the {\em kagom\'e} lattice are studied for the exactly solvable
infinite-component spin-vector model, $D \to \infty$.
In this limit, the critical coupling of fluctuations dies out and the critical
behavior simplifies, but the effect of would be Goldstone modes preventing
ordering at any nonzero temperature is properly accounted for.
In contrast to conventional two-dimensional magnets with continuous symmetry
showing extended short-range order at distances smaller than the
correlation length, $r \lsim \xi_c \propto \exp(T^*/T)$,  correlations in the {\em kagom\'e}-lattice
model decay already at the scale of the lattice spacing due to
the strong degeneracy of the ground state characterized by a macroscopic
number of strongly fluctuating local degrees of freedom.  
At low temperatures, spin CF's decay as $\langle {\bf S}_0 {\bf S}_r \rangle
\propto 1/r^2$ in the range $a_0 \ll r \ll \xi_c \propto T^{-1/2}$, where $a_0$ is
the lattice spacing.
Analytical results for
the principal thermodynamic quantities in our model are in fairly good quantitative 
agreement with the Monte Carlo simulations for the classical Heisenberg model, $D=3$. 
The neutron-scattering cross section has its maxima beyond the first Brillouin
zone; at $T\to 0$ it becomes nonanalytic but does not diverge at any ${\bf q}$.
\smallskip
\begin{flushleft}
PACS number(s):  75.50.Ee, 75.40.Cx
\end{flushleft}
} 
} 
\maketitle

\section{Introduction}

Classical antiferromagnets on {\em kagom\'e} and pyrochlore lattices built
of corner-sharing triangles and tetrahedra, respectively, are examples of
frustrated systems which cannot order because of the high degeneracy of their
ground state \cite{reibershi91} and ensuing large fluctuations.
Monte Carlo (MC) simulations for {\em kagom\'e} \cite{chaholshe92} and pyrochlore \cite{rei92}
lattices with the nearest-neighbor (nn) interaction $J$ show a smooth temperature dependence of the heat
capacity, $C(T)$, in the entire temperature range.
The spin correlation functions (CF's) of both models show only weak short-range
order at $T\lsim J$ and decay at distances of the order of several
lattice spacings $a_0$.
This is in a striking contrast to the long-range order in common
three-dimensional magnets and the extended short-range order [strong
correlation at the distances $r \lsim \xi_c \propto \exp(T^*/T)$ with $T^* \sim J$] in common
two-dimensional magnets.    

Spin-wave calculations starting from one of the ordered states of the {\em kagom\'e}
lattice \cite{harkalber92} (see, also, Ref. \onlinecite{zenels90} for the quantum case) yield a twofold degenerate Goldstone mode, as well
as a zero-energy mode for all values of the wave vector $q$ in the Brillouin
zone, the latter reflecting the instability of the ground states. 
Mean-field approximation (MFA) at elevated temperatures for both {\em kagom\'e} and
pyrochlore lattices \cite{reibershi91} reflects the same behavior.
The maximal eigenvalues of the Fourier-transformed exchange interaction matrix
are $q$ independent for both models (and twofold degenerate for pyrochlore).
Thus the system cannot choose the ordering wave vector at the mean-field
transition temperature, $T_c^{\rm MFA}$, which in fact shows that there is no
phase transition at this temperature because of fluctuations.
This results in the smooth temperature dependence of the thermodynamic
quantities \cite{chaholshe92,rei92} and in the diffuse magnetic neutron scattering. \cite{rei92ga}

The degeneracy of the ground state of these models can be lifted by
small perturbations, such as dipole-dipole interactions, lattice distortions,
next-nearest-neighbor (nnn) or long-range interactions, and quantum effects.
This may be a reason why pyrochlore antiferromagnets usually freeze into a  spin-glass state with lowering 
temperature. \cite{ram94,schram96}
Theoretically, the most transparent way to lift the degeneracy is to include nnn interactions in the Hamiltonian. 
\cite{reibershi91}
Experiment and MC simulations on pyrochlores \cite{reigrebjo92} show ordering with an unusual critical 
behavior ($\beta \approx 0.18$) in this case. 
According to the spin-wave results of Ref. \onlinecite{pichthes} for the {\em kagom\'e} lattice, at low temperatures 
dipole-dipole interactions favor the planar ${\bf q}=0$ phase which is characterized by the same ordering pattern in 
each of the elementary triangles.

A more subtle mechanism for lifting the degeneracy and selection of definite ordering patterns is the nonlinear 
interaction of spin waves for classical systems at very low temperatures, typically $T\lsim 0.01 J$.
For the {\em kagom\'e} lattice, nonlinear effects (thermal fluctuations) favor the coplanar spin configuration with the 
$\sqrt{3} \times \sqrt{3}$ short-range order in the case of the Heisenberg model,
$D=3$, as was suggested by the results of MC simulations \cite{chaholshe92,husrut92,reiber93} and high-temperature series expansions.
 \cite{harkalber92}
Extension of the $\sqrt{3} \times \sqrt{3}$ short-range order into the true long-range order in the limit $T\to 0$ is,
however, hampered by formation of chiral domain walls which cost no energy but provide a gain in entropy at low
concentrations. \cite{reiber93} 
The configuration selection at low temperatures only occurs if the number of spin components $D$ is
low enough.
So, the early MC simulations of Ref.\onlinecite{husrut92} for the {\em kagom\'e} antiferromagnet showed selection of
a coplanar state for $D=3$, but no such selection for $D\geq 4$.  
For the pyrochlore lattice, early simulations showed the selection of the collinear spin ordering for the Heisenberg 
model  at low temperatures, \cite{reigrebjo92} although according to the recent results of Ref. \onlinecite{moecha98} 
this happens only for the plane rotator model, $D=2$, and not for higher spin dimensionalities.
The above results are in accord with general criterion for selection of ordered states as a function of spin and space 
dimensions for corner sharing objects, which was formulated in Ref. \onlinecite{moecha98}.
Quantum fluctuations were shown to stabilize the $\sqrt{3} \times \sqrt{3}$ phase for $S\gg 1$, \cite{chub92} but they should
destroy ordering for low spin values $S$. \cite{zenels90,sac92,leuels93,lecetal97}
One of the possible mechanisms for that is tunneling of the weatherwane (hard hexagon) mode in the 
$\sqrt{3} \times \sqrt{3}$ structure. \cite{delhen9293}

It should be stressed, however, that the subtle effects quoted above can be easily overwhelmed by more trivial 
and robust ones, and they are much easier to observe in simulations than in experiment.
The first task of the theory is thus to describe the principal features of classical spin models on frustrating lattices, 
as, e.g., a smooth variation of the thermodynamic quantities in the whole temperature range.
The simplest approach, the MFA, is clearly inapplicable in this case, whereas
the more powerful tools of the theory of critical phenomena, such as the renormalization group, seem to have not been yet applied to these lattices.  

The ``next simplest'' approximation for classical spin systems, which follows the MFA, consists in generalizing the Heisenberg Hamiltonian for the $D$-component spin vectors:
\cite{sta68prl,sta74}
%
\begin{equation}\label{dham}
{\cal H} = -\bbox{\rm H \cdot}\sum_{r}{\bf s}_r -
 \frac{1}{2}\sum_{rr'}J_{rr'}{\bf s}_r \bbox{ \cdot \rm s}_{r'}  ,  \qquad |{\bf s}_r|=1
\end{equation}
and taking the limit $D\to\infty$.
In this limit the problem becomes exactly solvable for all lattice dimensionalities, $d$, and the partition function of the system coincides \cite{sta68pr} with that of the spherical model. \cite{berkac52,joy72pt}
The $D=\infty$ model possesses, however, a number of important advantages with respect to the spherical one. 
(i) The $1/D$ expansion is possible, \cite{abe7273,abehik7377,okamas78} including the case of low-dimensional systems. \cite{gar94jsp,gar96jsp}
The calculations can be done conveniently in the framework of the diagram technique for classical spin systems. \cite{garlut84d,gar94jsp,gar96prb}
(ii) Inclusion of anisotropic terms in Eq. (\ref{dham}) is possible, too,
which allows us to describe ordering in low dimensions, including thin films \cite{gar96jpal} and domain walls. \cite{gar96jpa}
(iii) In spatially inhomogeneous cases the $D=\infty$ model yields physically correct results, in contrast to the spherical model failing on the global spin constraint. \cite{barfis73}
(iv) Below the Curie temperature $T_c$ or in a magnetic field, the $D=\infty$
model describes both transverse and longitudinal CF's (Ref. \onlinecite{gar97zpb}) that differ from each other, in contrast to the single CF in the spherical model.

The $D=\infty$ model properly accounts for the profound role played, especially in low dimensions, by the Goldstone or would be Goldstone modes.
At the same time,  the less significant effects of the critical fluctuation coupling leading, e.g., to the quantitatively different nonclassical critical indices, die out in the limit $D\to\infty$.
Thus this model is a relatively simple yet a powerful tool for classical spin systems.
It should not be mixed up with the $N$-flavor generalization of the quantum $S=1/2$ model 
\cite{auearo8888}  in the limit $N\to\infty$, including its $1/N$ expansion. \cite{sac92,timgirhen98}
The $N$-component nonlinear $\sigma$-model (see, e.g.,
Refs. \onlinecite{chahalnel8889}, as well as Ref. \onlinecite{chubsacye94},
and \onlinecite{chubsta98} for the $1/N$ expansion) is a quantum extension of  Eq. (\ref{dham}) in the long-wavelength region at low temperatures.
Effective free energies for the $n$-component order parameter appear, instead of Eq. (\ref{dham}), in conventional theories of critical phenomena. 
Using them for the $1/n$ expansion (see, e.g., Ref. \onlinecite{ma73}) is a matter of taste.
While yielding the same results for the critical indices as the lattice-based $1/D$ expansion,
\cite{abe7273,abehik7377,okamas78} it misses the absolute values of the nonuniversal quantities.
The same comment also applies to the spatially inhomogeneous systems in the
limit $D=n=\infty$, such as semi-infinite ferromagnets (cf. Refs. \onlinecite{bramoo7777} and \onlinecite{gar98pre}).

In this article the solution for the isotropic antiferromagnetic infinite-component spin-vector model on the {\em kagom\'e} lattice will be given.
The qualitatively similar results for the pyrochlore lattice will be presented in a subsequent communication.
As long as the system studied is homogeneous, isotropic, and in zero magnetic field, 
the standard spherical model \cite{berkac52,joy72pt} can be applied, too.
Such an approach for nonordering frustrated three-dimensional systems has been advocated in Ref. \onlinecite{pimras90}.
We prefer, however, to use the more general framework.

The rest of this article is organized as follows.
In Sec. \ref{secStructure} the structure of the {\em kagom\'e} lattice and its collective spin variables are described.
In Sec. \ref{secDinfinity} the formalism of the $D=\infty$ model is tailored for the {\em kagom\'e} lattice. 
The diagrams of the classical spin diagram technique that do not disappear in the limit $D\to\infty$ are summed up.
The general analytical expressions for the thermodynamic functions and spin CF's for all temperatures are obtained.
In Sec. \ref{secThermod} the thermodynamic quantities of the {\em kagom\'e}
antiferromagnet (AFM) are calculated and
compared with MC simulation results in the whole temperature range.
In Sec. \ref{secCF's} the real space correlation functions are computed.
In Sec. \ref{secNeutron} the neutron
scattering cross section is worked out.
In Sec. \ref{secDiscussion} possible improvements of the present approach, such as the $1/D$ expansion, are discussed.

\section{Lattice structure and the Hamiltonian}
\label{secStructure}

\begin{figure}[t]
\unitlength1cm
\begin{picture}(11,7)
\centerline{\epsfig{file=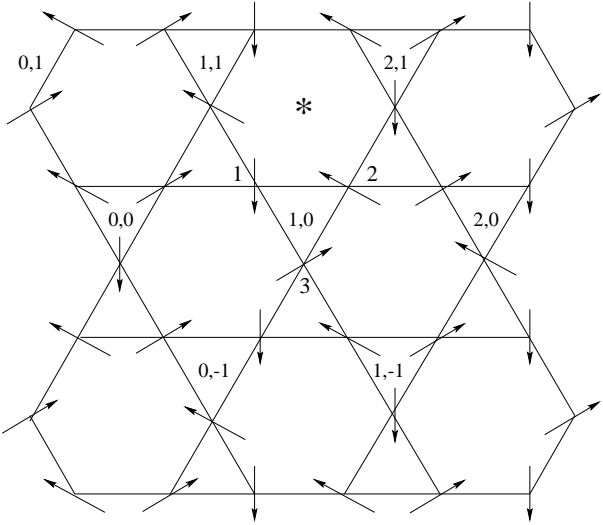,angle=0,width=8cm}}
\end{picture}
\caption{ \label{kag_str} 
Structure of the {\em kagom\'e} lattice.
The elementary triangles are labeled by the pairs of numbers $n_u, n_v$
according to Eq. (\protect\ref{transl}), and the sites on triangles (the
sublattices) are labeled by $l=1,2,3$.
The configuration shown corresponds to the coplanar $\protect\sqrt{3} \times \protect\sqrt{3}$
structure characterized by the ordering wave vector 
${\protect\bf q}_{\protect\sqrt{3}}$ given by  Eq. (\protect\ref{defqsq3}).
}
\end{figure}

The {\em kagom\'e} lattice shown in Fig. \ref{kag_str} consists of corner-sharing
triangles.
 Each node of the corresponding Bravais lattice (i.e.,  each elementary
triangle in Fig. \ref{kag_str}) is numbered by $i,j=1,\ldots, N$.
Each site of the elementary triangle is labeled by the index $l=1,2,3$.
It is convenient to use the dimensionless units in which the interatomic
distance equals 1 and hence the lattice period equals 2.
The triangles numbered by $i,j=1,\ldots, N$ can be obtained from each other by
the translations
%
\begin{equation}\label{transl}
{\bf r}_j^l =  {\bf r}_i^l + n_u 2{\bf u} + n_v 2{\bf v} ,
\end{equation}
where $ {\bf r}_i^l$ is the position of a site on the lattice, $n_u$ and $n_v$
are integers, $2{\bf u}$  and $2{\bf v}$ are the elementary translation vectors (lattice periods), and
%
\begin{equation}\label{defuv}
{\bf u} = ( 1, 0 ), 
\qquad {\bf v} = ( -1/2,  \sqrt{3}/2 ) .
\end{equation}

One of the most symmetric phases of the {\em kagom\'e} AFM is the so-called 
$\sqrt{3} \times \sqrt{3}$ phase which is shown in
Fig. \ref{kag_str}.
This coplanar phase can be described by the complex ``spin'' 
$\tilde s_{\bf r} \equiv  s_{\bf r}^x + i  s_{\bf r}^y =
 \exp(i  {\bf q}_{\sqrt{3}} \bbox{\rm \cdot r} + i\phi_0 )$, where the ordering wave
vector ${\bf q}_{\sqrt{3}}$ can be written in three equivalent forms:
%
\begin{eqnarray}\label{defqsq3}
&&
{\bf q}_{\sqrt{3}} = - \frac{ 2\pi}{ 3 } \times \left\{ 
\begin{array}{l}
 {\bf u }\\
 {\bf v }\\
 {\bf w } ,
\end{array}
\right.
\end{eqnarray}
where ${\bf w} = ( -1/2, - \sqrt{3}/2 )$.
In this phase spins rotate by $-240^\circ=120^\circ$ as ${\bf r}$ changes by the
lattice period 2 in each of the directions ${\bf u }$, ${\bf v }$, and
${\bf w }$ making the angle $120^\circ$ with each other.
Another realization of the $\sqrt{3} \times \sqrt{3}$ phase, in which spins
rotate by $-120^\circ$, is described by  ${\bf q}_{\sqrt{3}}$
with positive sign.
In addition, the $\sqrt{3} \times \sqrt{3}$ phase can be described by appropriate
combinations of different forms of ${\bf q}_{\sqrt{3}}$ given above.

\begin{figure}[t]
\unitlength1cm
\begin{picture}(11,8)
\centerline{\epsfig{file=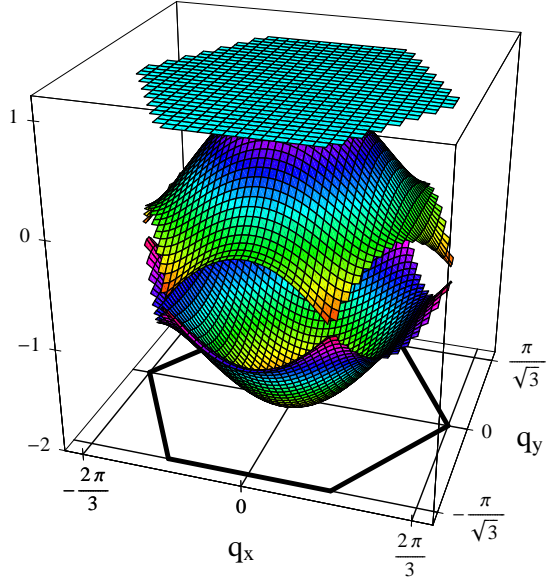,angle=0,width=9cm}}
\end{picture}
\caption{ \label{kag_mod} 
Reduced eigenvalues of the interaction matrix, $ \nu_n({\bf q}) = \tilde V_{\bf q}^n/(2J)$ of Eq. (\protect\ref{dhamdiag}), plotted over the Brillouin zone.
}
\end{figure}

To facilitate the diagram summation in the next section, it is convenient to put the Hamiltonian (\ref{dham}) into a diagonal form.
First, one goes to the Fourier representation according to
%
\begin{equation}\label{fourier}
{\bf s}_{\bf q}^l = \sum_i {\bf s}_i^l e^{-i \bbox{\rm q\cdot r}_i^l}, 
\qquad {\bf s}_i^l = \frac 1N \sum_{\bf q} {\bf s}_{\bf q}^l e^{i \bbox{\rm q\cdot r}_i^l}, 
\end{equation}
where the wave vector ${\bf q}$ belongs to the hexagonal Brillouin zone
specified by the corners $(\pm \pi/3, \pm \pi/\sqrt{3})$ and $(\pm
2\pi/3,0)$ (see Fig. \ref{kag_mod}).
The Fourier-transformed Hamiltonian reads
%
\begin{equation}\label{dhamq}
{\cal H} = \frac{1}{2N} \sum_{ll'\bf q} V_{\bf q}^{ll'} {\bf s}_{\bf q}^l 
\bbox{ \cdot \rm s}_{-{\bf q}}^{l'} , 
\end{equation}
where the interaction matrix is given by
%
\begin{equation}\label{Vq}
\hat V_{\bf q} = 2J
\left( \begin{array}{ccc}
0 & a & b \\
a & 0 & c \\
b & c & 0 
\end{array} \right) ,
\qquad 
\begin{array}{l}
a \equiv \cos( {\bf u \cdot q}) \\
b \equiv \cos( {\bf v \cdot q}) \\
c \equiv \cos( {\bf w \cdot q}). \\
\end{array}
\end{equation}

At the second stage, the Hamiltonian (\ref{dhamq}) is finally diagonalized to the form
%
\begin{equation}\label{dhamdiag}
{\cal H} =  -\frac{1}{2N} \sum_{n\bf q} \tilde V_{\bf q}^n 
\bbox{\sigma}_{\bf q}^n \bbox{\cdot \sigma}_{-\bf q}^n , 
\end{equation}
where $ \tilde V_{\bf q}^n = 2J \nu_n({\bf q})$ are the eigenvalues of the matrix $V_{\bf q}^{ll'} $ taken with the negative sign,
%
\begin{equation}\label{nudef}
\nu_1 = 1, \qquad \nu_{2,3} = (\pm \sqrt{1+8abc} - 1 )/2.
\end{equation}
The diagonalizing transformation has the explicit form 
%
\begin{equation}\label{Vtrans}
U_{nl}^{-1}({\bf q}) V^{ll'}_{\bf q} U_{l'n'}({\bf q}) = \tilde V^n_{\bf q} \delta_{nn'} ,
\end{equation}
where the summation over the repeated indices is implied and $\hat U$ is the real unitary matrix,
$\hat U^{-1} = \hat U^T$, i.e., $U_{nl}^{-1} = U_{ln}$.
The columns of the matrix $\hat U$ are the three normalized
eigenvectors $U_n = (U_{1n},U_{2n},U_{3n})$ of the interaction matrix $\hat V$:
%
\begin{eqnarray}\label{eigenvec}
&&
U_n  = ( ac-b\nu_n, \; ab-c\nu_n, \; \nu_n^2-a^2 )/\sqrt{Q_n},
\nonumber\\
&&
Q_n = (\nu_n^2-a^2)^2 + (ab-c\nu_n)^2 + (ac-b\nu_n)^2.
\end{eqnarray}
The eigenvector $U_1$ corresponding to the dispersionless eigenvalue $\nu_1=1$
can be represented in the unnormalized form as
%
\begin{equation}\label{eigenvec1}
U_1  = [ \sin( \bbox{\rm w \cdot q}),  \; \sin( \bbox{\rm v \cdot q}),  \; \sin( \bbox{\rm u \cdot q}) ].
\end{equation}
The normalized eigenvectors satisfy the requirements of orthogonality 
and completeness, respectively,
%
\begin{equation}\label{orthcomp}
U_{ln}({\bf q}) U_{ln'}({\bf q}) = \delta_{nn'},
\qquad U_{ln}({\bf q}) U_{l'n}({\bf q}) = \delta_{ll'} .
\end{equation}
The Fourier components of the spins ${\bf s}_{\bf q}^l$ and the 
collective spin variables $\bbox{\sigma}_{\bf q}^n$ are related by
%
\begin{equation}\label{s-sigma}
{\bf s}_{\bf q}^l = U_{l n}({\bf q}) \bbox{\sigma}_{\bf q}^n, 
\qquad \bbox{\sigma}_{\bf q}^n = {\bf s}_{\bf q}^l U_{l n}({\bf q}).
\end{equation}

The largest dispersionless eigenvalue $\nu_1$ of the interaction matrix [see Eq. (\ref{nudef})] manifests frustration in the system which precludes an extended short-range order even in the limit $T\to 0$.
Independence of $\nu_1$ of ${\bf q}$ signals that 1/3 of all spin degrees of freedom are local and can rotate freely.
The other two eigenvalues satisfy
%
\begin{equation}\label{nu23qsm}
\nu_2 ({\bf q}) \cong 1 -q^2/2, \qquad \nu_3 ({\bf q}) \cong -2 + q^2/2
\end{equation}
at small wave vectors, $q^2 \equiv q_x^2 + q_y^2 \ll 1$.
The eigenvalue $\nu_2$ which becomes degenerate with $\nu_1$ in the limit $q\to 0$ is related, as we shall see below, to the usual would be Goldstone mode destroying the long-range order in low-dimensional magnets with a continuous symmetry.
The eigenvalue $\nu_3$ is positioned,  in the long-wavelength region, much lower than the first two, and it is tempting to call it the ``optical'' eigenvalue. 
In fact, however, the eigenvalues of the interaction matrix are {\em not} the same as the {\em normal modes} of the system that appear in the dynamics.
Whereas $\nu_1$ gives rise to the zero-energy spin-wave branch corresponding to the absence of a restoring force for small deviations from one of degenerate ground states, $\nu_2$ and $\nu_3$ hybridize to the double-degenerate Goldstone mode with energy $\propto q$ at small wave vectors, as in conventional antiferromagnets. \cite{zenels90,harkalber92}
 With increasing $q$ the eigenvalue $\nu_2$ decreases whereas $\nu_3$ increases; at the corners of the Brillouin zone they become degenerate: $\nu_2=\nu_3=-1/2$.
The ${\bf q}$ dependences of the eigenvalues $\nu_n$ over the whole Brillouin zone is shown in Fig. \ref{kag_mod}.

In contrast to the smooth behavior of $\nu_n$, the diagonalizing matrix $\hat U$ composed of the eigenvectors $U_n$ [see Eq. (\ref{eigenvec})] has much more complicated structure as function of ${\bf q}$.
This results in the intricate behavior of the spin CF's and neutron-scattering
cross sections at low temperatures, which will be considered in Sec. \ref{secCF's}.
Here we only show the nonanalytic limiting form of $\hat U$ at small wave vectors,
%
\begin{equation}\label{Uqsm}
\hat U_{\bf q} \cong \frac{ 1 }{ \sqrt{6} }
\left( \begin{array}{ccc}
-n_x - \sqrt{3} n_y    &    -n_y + \sqrt{3} n_x    &    \sqrt{2}   \\
-n_x + \sqrt{3} n_y     &    -n_y - \sqrt{3} n_x   &    \sqrt{2}   \\
2 n_x                   &    2 n_y                 &    \sqrt{2} 
\end{array} \right) ,
\end{equation}
where ${\bf n} \equiv  {\bf q}/q$.

In the next section the equations describing spin correlation functions of the
classical {\em kagom\'e} antiferromagnet in the large-$D$ limit will be obtained with
the help of the classical spin diagram technique.
The readers who are not interested in details can skip to Eq. (\ref{defcfsig})
or directly to Sec. \ref{secThermod}.

\section{Classical spin diagram technique and the large-$D$ limit}
\label{secDinfinity}

\begin{figure}
\unitlength1cm
\begin{picture}(11,4)
\centerline{\epsfig{file=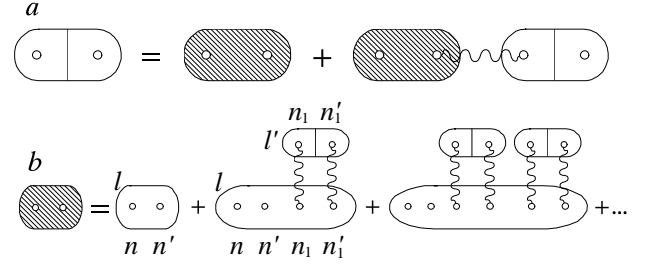,angle=0,width=16cm}}
\end{picture}
%
\caption{ \label{kag_scga}
Self-consistent Gaussian approximation (SCGA) for classical spin 
systems in the nonordered state.
($a$) the Dyson equation for the spin correlation function; 
($b$) the block summation for the renormalized pair cumulant spin averages.
}
\end{figure}

The exact equations for spin correlation functions in the limit $D\to\infty$, as well as the $1/D$ corrections,
can be the most conveniently obtained with the help of the classical spin diagram technique.
\cite{garlut84d,gar94jsp,gar96prb} 
A perturbative expansion of the thermal average of any quantity ${\cal A}$ 
characterizing a classical spin system 
(e.g., ${\cal A} = s_{zi}$ --- the $z$ spin component on the lattice site $i$)
can be obtained by rewriting Eq. (\ref{dham}) as 
${\cal H} ={\cal H}_0 + {\cal H}_{\rm int}$, 
where ${\cal H}_0$ is, e.g., the mean-field Hamiltonian, 
and expanding the expression 
%
\begin{equation}\label{statavr}
\langle{\cal A}\rangle  = \frac{1}{{\cal Z}} \int\prod_{j=1}^N d{\bf s}_j
{\cal A} \exp(-\beta {\cal H}), \qquad  |{\bf s}_j|=1 ,
\end{equation}
where $\beta=1/T$, in powers of ${\cal H}_{\rm int}$. 
The integration in Eq. (\ref{statavr}) is carried out with respect to the 
orientations of the $D$-dimensional unit vectors ${\bf s}_j$ on each of the lattice sites.
Averages of various spin-vector components 
on various lattice sites with the Hamiltonian ${\cal H}_0$ can be 
expressed through spin cumulants, or semi-invariants, which will be 
considered below, in the following way:
%
\begin{eqnarray}\label{siteavr}
&& 
\langle s_{\alpha i}\rangle _0  =  \Lambda_\alpha, \nonumber   \\
&&
\langle s_{\alpha i}s_{\beta j}\rangle _0 = \Lambda_{\alpha\beta} \delta_{ij} 
+ \Lambda_\alpha \Lambda_\beta,         \\
&& 
 \langle s_{\alpha i}s_{\beta j}s_{\gamma k}\rangle _0  =  
 \Lambda_{\alpha\beta\gamma} \delta_{ijk} 
+ \Lambda_{\alpha\beta}\Lambda_\gamma \delta_{ij} \nonumber  \\
&&  \qquad \qquad \qquad 
+\, \Lambda_{\beta\gamma}\Lambda_\alpha \delta_{jk}
+ \Lambda_{\gamma\alpha}\Lambda_\beta \delta_{ki}
+ \Lambda_\alpha \Lambda_\beta \Lambda_\gamma ,        \nonumber
\end{eqnarray}
etc., where $\delta_{ij}$, $\delta_{ijk}$, etc., are the site Kronecker 
symbols equal to 1 for all site indices coinciding with each other and to 
zero in all other cases. 
For the one-site averages 
($i=j=k=\ldots$) Eq. (\ref{siteavr}) reduces to the well-known 
representation of moments through semi-invariants, generalized for 
a multicomponent case. 
In the graphical language (see Fig. \ref{kag_scga}) the decomposition 
(\ref{siteavr}) corresponds to all possible groupings of small circles 
(spin components) into oval blocks (cumulant averages). 
The circles 
coming from ${\cal H}_{\rm int}$ (the ``inner'' circles) 
are connected pairwise by the wavy interaction lines representing   
$\beta J_{ij}$. 
In diagram expressions, summations over site indices 
$i$ and component indices $\alpha$ of inner circles are carried out. 
One should not take into account disconnected (unlinked) diagrams [i.e., those 
containing disconnected parts with no ``outer'' circles belonging to 
${\cal A}$ in Eq. (\ref{statavr})], since these diagrams are compensated for 
by the expansion of the partition function ${\cal Z}$ in the denominator 
of Eq. (\ref{statavr}). 
Consideration of combinatorial numbers shows 
that each diagram contains the factor $1/n_s$, where $n_s$ is the number 
of symmetry group elements of a diagram [see, e.g., the factor $1/2!$ in Eq. (\ref{L20}) below].   
The symmetry operations do not concern outer circles,
which serve as a distinguishable ``root'' to build up more complicated (e.g.,  
renormalized) diagrams. 
For spatially homogeneous systems, it is more 
convenient to use the Fourier representation and to calculate integrals 
over the wave vectors in the Brillouin zone rather than lattice sums. 
As due to the Kronecker symbols in Eq. (\ref{siteavr}) lattice sums 
are subject to the constraint that the coordinates of the circles 
belonging to the same block coincide with each other, the sum of wave vectors coming to or going out of any block along 
interaction lines is zero. 
So, for our model the pair cumulant average of the Fourier components defined
by Eq. (\ref{fourier}) reads
%
\begin{equation}\label{Fourcum}
\langle s_{\alpha \bf q}^l s_{\beta {\bf q}'}^{l'} \rangle_{0,\rm cum} =
\Lambda_{\alpha\beta} N \delta_{ {\bf q}', -{\bf q} } \delta_{ll'},
\end{equation}
where $\delta_{ll'}$ is the sublattice Kronecker symbol.
The cumulant spin averages in Eq. (\ref{siteavr}) 
can be obtained by differentiating the generating function $\Lambda(\xi)$ 
over appropriate components of the dimensionless field 
$\bbox{\xi}\equiv \beta{\bf H}$: \cite{garlut84d}
%
\begin{eqnarray}\label{defcum}
&&
\Lambda_{\alpha_1\alpha_2\ldots\alpha_p}(\bbox{\xi})=
\frac{\partial ^p\Lambda(\xi)}{\partial \xi_{\alpha_1} 
\partial \xi_{\alpha_2}\ldots\partial \xi_{\alpha_p}}, \nonumber\\ 
&&
\Lambda(\xi)=\ln {\cal Z}_0(\xi),
\end{eqnarray}
where $\xi \equiv |\bbox{\xi}|$,
%
\begin{equation}\label{partfunc}
{\cal Z}_0(\xi)={\rm const}\times \xi^{-(D/2-1)} I_{D/2-1}(\xi)
\end{equation}
is the partition function of a $D$-component classical spin, and 
$I_\nu(\xi)$ is the modified Bessel function. 
For the two lowest-order cumulants the differentiation in Eq. (\ref{defcum}) 
leads to the following expressions:
%
\begin{eqnarray}\label{cum}
&&
\Lambda_\alpha(\bbox{\xi}) = B(\xi)\,\xi_\alpha/\xi ,     \nonumber  \\
&&
\Lambda_{\alpha\beta}(\bbox{\xi}) 
= \frac{B(\xi)}{\xi}
\left( \delta_{\alpha\beta} - \frac{\xi_\alpha \xi_\beta}{\xi^2} \right)
+ B'(\xi) \frac{\xi_\alpha \xi_\beta}{\xi^2} .
\end{eqnarray}
where $\delta_{\alpha\beta}$ is the  Kronecker symbol for spin components, 
%
\begin{equation}\label{defb}
B(\xi) = d\Lambda(\xi)/d\xi = I_{D/2}(\xi)/I_{D/2-1}(\xi)
\end{equation}
is the Langevin function of $D$-component classical spins, and
$B'(\xi)=dB/d\xi$ (see the details in Ref. \onlinecite{gar96prb}).
If ${\cal H}_0 =0$, as is the case for our model in the absence of a magnetic field, the
pair spin cumulant in Eq. (\ref{cum}) simplifies to the obvious form
%
\begin{equation}\label{cum2free}
\Lambda_{\alpha\beta}(0) = \Lambda_{\alpha\alpha}(0) \delta_{\alpha\beta},
\qquad \Lambda_{\alpha\alpha}(0) = 1 /D.
\end{equation}

As was shown in Ref. \onlinecite{garlut84d} (see also Ref. \onlinecite{gar94jsp})
the limit $D\to\infty$ for the spin-vector model is completely described by the 
self-consistent Gaussian approximation (SCGA), since all diagrams not accounted
for by the SCGA vanish in this limit.
The SCGA consists in taking into 
account {\em pair} correlations of the molecular field acting on a given 
spin from its neighbors, which implies a Gaussian statistics of 
molecular field fluctuations. 
The appropriate diagram sequence for the nonordered state, 
$\langle s_z \rangle = 0$, is 
represented in Fig. \ref{kag_scga}.
Its analytical form for the square lattice model is given in 
Ref. \onlinecite{gar94jsp}.
In a magnetic field or below $T_c$ in the ordering models the average spin 
polarization $\langle s_z \rangle \ne 0$ appears.
The additional diagrams and corresponding analytical expressions can be found in 
Refs. \onlinecite{garlut84d,gar96prb}, and \onlinecite{gar96jsp}.
The SCGA equations in the spatially inhomogeneous case and their large-$D$ limit
have been derived (and applied to domain walls) in Ref. \onlinecite{gar96jpa}.   

In all cases above, the SCGA equations have been written for {\em
diagonal} Hamiltonians describing the simplest one-sublattice magnets.
For nondiagonal Hamiltonians, such as Eq. (\ref{dhamq}), the matrix
interaction lines, here  $-\beta V_{\bf q}^{ll'}$, complicate the formalism.
Simplification can be achieved by using the diagonalized Hamiltonian, for our
model, Eq. (\ref{dhamdiag}).
For the latter, however, the $\sigma$ counterparts of the one-site cumulant
averages do not have a transparent meaning anymore since $\sigma$ is a
combination of spins on different sites and sublattices.
Thus the $\sigma$ cumulants should be specially worked out as follows.
The pair $\sigma$ cumulant (which is in our model explicitly diagonal in the
spin-component indices $\alpha,\beta$) can be rewritten in terms of the
initial spin variables as
%
\begin{equation}\label{cusipair}
\langle \sigma_{\alpha \bf q}^n \sigma_{\alpha {\bf q}'}^{n'} \rangle_{0,\rm cum} =
U_{ln}({\bf q}) U_{l'n'}({\bf q}')
\langle s_{\alpha \bf q}^l s_{\alpha {\bf q}'}^{l'} \rangle_{0,\rm cum}.
\end{equation}
With the  use of Eq. (\ref{Fourcum}) and the first of the relations
(\ref{orthcomp}), this can be simplified to the final form
%
\begin{equation}\label{cusipair2}
\langle \sigma_{\alpha \bf q}^n \sigma_{\alpha {\bf q}'}^{n'} \rangle_{0,\rm cum} =
\Lambda_{\alpha\alpha} N \delta_{ {\bf q}', -{\bf q} } \delta_{nn'}.
\end{equation}
Now, with the help of the results just obtained, the second diagram in the sum
in Fig. \ref{kag_scga}$b$ can be written in the analytical form
%
\begin{equation}\label{cusifour}
A_2 = U_{l n}({\bf q}) U_{l n'}({\bf q}) \Lambda_{\alpha\alpha\beta\beta}
L_{\beta l} ,
\end{equation}
where the summation over the spin-component index $\beta$ is implied and the
quantity $L_{\beta l}$ in the lowest (the second) order of the perturbation
theory is  given by
%
\begin{eqnarray}\label{L20}
&&
L_{\beta l}^{(2)} = \frac { 1 }{ 2! }  \Lambda_{\beta\beta} \frac 1N \sum_{l' \bf p} 
\sum_{n_1} U_{l n_1}({\bf p}) U_{l' n_1}({\bf p}) \beta \tilde V_{\bf p}^{n_1} 
\nonumber\\
&&
\qquad {} \times
\sum_{n'_1} U_{l n'_1}({\bf p}) U_{l' n'_1}({\bf p}) \beta \tilde V_{\bf p}^{n'_1} .
\end{eqnarray}
Here $\beta=1/T$ in front of $ \tilde V$ cannot be confused with the spin
component index $\beta$.
This expression can be simplified in two ways. 
First, one can perform the sum over the index
$l'$ and use the first of Eqs. (\ref{orthcomp}), which leads to
%
\begin{equation}\label{L2n}
L_{\beta l}^{(2)} = \frac { 1 }{ 2! }  \Lambda_{\beta\beta} \frac 1N
\sum_{n_1 \bf p} ( \beta \tilde V_{\bf p}^{n_1} )^2 U_{l n_1}^2({\bf p}).
\end{equation}
Second, inverting the transformation (\ref{Vtrans}) one can write
%
\begin{equation}\label{L2l}
L_{\beta l}^{(2)} = \frac { 1 }{ 2! }  \Lambda_{\beta\beta} \frac 1N
\sum_{l' \bf p}  ( \beta V_{\bf p}^{ll'} )^2 .
\end{equation}
Taking into account the explicit form of $V_{\bf p}^{ll'}$ given by
Eq. (\ref{Vq}), one can see that after the integration over the wave vector
${\bf p}$ expression (\ref{L2l}) becomes independent of the sublattice 
index $l$.
After this observation one can symmetrize Eq. (\ref{L2n}) with respect to $l$.
This leads to the vanishing of the diagonalization matrix by virtue of the
first of Eqs. (\ref{orthcomp}) and to the appearance of the factor 1/3.
Now the summation over $l$ in Eq. (\ref{cusifour}) simplifies,  and the
diagonalization matrices convert, again, to $\delta_{nn'}$.
The result in the second order of the perturbation theory has the form
%
\begin{equation}\label{cusifour2}
A_2 =  \Lambda_{\alpha\alpha\beta\beta} L_\beta^{(2)} ,
\qquad L_\beta^{(2)} = \frac{  \Lambda_{\beta\beta} }{ 3 \cdot 2! } \frac 1N \sum_{n_1 \bf p} 
( \beta \tilde V_{\bf p}^{n_1} )^2
\end{equation}
($\delta_{nn'}$ has been omitted) and it is independent of the eigenvalue
index $n$ and of the wave vector ${\bf q}$.

The mechanism of the simplification of diagram expressions demonstrated above
 can be shown to work for whatever complicated diagrams.
In all cases oval blocks represent cumulant spin averages 
$\Lambda_{\alpha_1\alpha_2 \ldots \alpha_p}$, as in the original,
nondiagonalized, version of the classical spin diagram technique.
In all the elements connected to a given block summation over the eigenvalue
indices $n$ is carried out.
The diagonalizing matrix $\hat U$ disappears completely if correlation
functions for the $\sigma$ variables,
%
\begin{equation}\label{defcfsig}
\sigma^n({\bf q}) = \frac DN 
\langle \sigma_{\alpha \bf q}^n \sigma_{\alpha, {-\bf q}}^n \rangle 
=\frac 1N \langle \bbox{\sigma}_{ \bf q}^n \bbox{\sigma}_{-\bf q}^n \rangle ,
\end{equation}
are considered. 
After the calculation of the latter the true spin CF's can be found from the
formula
%
\begin{equation}\label{defcfs}
s^{ll'}({\bf q}) = U_{l n}({\bf q}) U_{l' n}({\bf q})
\sigma^n({\bf q}) 
\end{equation}
following from Eq. (\ref{s-sigma}).
Note that $\sigma^n({\bf q})$ are eigenvalues of the correlation matrix
$s^{ll'}({\bf q}) $ and they describe independent linear responses to appropriate
wave-vector-dependent fields.
As can be seen from Eq. (\ref{defcfs}), the eigenvectors describing the
 ``normal modes'' of the  susceptibility are those of the
interaction matrix $V^{ll'}_{\bf q} $ in Eq. (\ref{dhamq}).

The analytical expression for the $\sigma$ CF in the SCGA, which satisfies the
Dyson equation shown in Fig. \ref{kag_scga}$a$, has the Ornstein-Zernike form
%
\begin{equation}\label{sigcfOZ}
\sigma^n({\bf q}) = \frac{ D \tilde\Lambda_{\alpha\alpha} }
{ 1 -  \tilde\Lambda_{\alpha\alpha} \beta \tilde V_{\bf q}^n } .
\end{equation}
This expression differs from that obtained by Reimers on the mean-field basis
\cite{rei92ga} by the replacement of the bare cumulant $\Lambda_{\alpha\alpha} =
1/D$ by its renormalized value $ \tilde\Lambda_{\alpha\alpha}$ determined by the diagram series
Fig. \ref{kag_scga}$b$. 
The summation of these diagrams is documented in the most detailed way by
Eqs. (3.16)--(3.19) of Ref.  \onlinecite{gar96prb}.
The result for $ \tilde\Lambda_{\alpha\alpha}$ is given by the second line of
Eq. (\ref{cum}) averaged over the {\em Gaussian} fluctuations of all components
of the molecular field $\bbox{\xi}$ with the dispersion defined by the
quantity $L_\alpha$.
In our model, fluctuations of different components of $\bbox{\xi}$ are 
independent from each other and of the same dispersion,  $L_\alpha=L$.  
Thus the quantity  $ \tilde\Lambda_{\alpha\beta}$
is diagonal and independent of $\alpha$.
In the large-$D$ limit the multiple Gaussian integral determining  $ \tilde\Lambda_{\alpha\alpha}$ 
is dominated by the stationary point and the result simplifies to \cite{gar94jsp}
%
\begin{equation}\label{Lamtil}
\tilde\Lambda_{\alpha\alpha} = \frac 2D \frac{ 1 }{ 1 + \sqrt{ 1 + 8L/D } } .
\end{equation}
Here the dispersion $L$ corresponding to the diagram series in
Fig. \ref{kag_scga} is given by the formula
%
\begin{equation}\label{L}
L = \frac{ \tilde \Lambda_{\alpha\alpha} }{ 3 \cdot 2! } \sum_n
 v_0\!\!\!\int\!\!\!\frac{d{\bf q}}{(2\pi)^d}
 \frac{ ( \beta \tilde V_{\bf q}^n )^2 }
{  1 -  \tilde\Lambda_{\alpha\alpha} \beta \tilde V_{\bf q}^n }
\end{equation}
generalizing Eq.~(\ref{cusifour2}).
Here, the summation $(1/N)\sum_{\bf q}\ldots$ is replaced by the integration over the
Brillouin zone, $v_0$ is the unit-cell volume, and $d$ is the spatial
dimensionality.
For the {\em kagom\'e} lattice we have $v_0 = 2\sqrt{3}$ and $d=2$.
The expression for $L$ can be simplified to
%
\begin{equation}\label{LbarP}
L = \frac{ \bar P - 1}{ 2\tilde\Lambda_{\alpha\alpha} }, 
\qquad \bar P \equiv \frac 13 \sum_n P_n ,
\end{equation}
where $P_n$ is the lattice Green function associated with the eigenvalue $n$:
%
\begin{equation}\label{Pndef}
P_n =  v_0\!\!\!\int\!\!\!\frac{d{\bf q}}{(2\pi)^d}
 \frac{ 1 }
{  1 -  \tilde\Lambda_{\alpha\alpha} \beta \tilde V_{\bf q}^n } .
\end{equation}
Now one can eliminate $L$ from Eqs. (\ref{Lamtil}) and (\ref{LbarP}), which
yields the basic equation of the large-$D$ model,
%
\begin{equation}\label{sphereq0}
D \tilde\Lambda_{\alpha\alpha} \bar P = 1 .
\end{equation}
This nonlinear equation determining $\tilde\Lambda_{\alpha\alpha}$ as
a function of temperature differs from those considered earlier 
\cite{garlut84d,gar94jsp,gar96jsp,gar96prb} by a more complicated form of
$\bar P$ reflecting the lattice structure.
The form of this equation is similar to that appearing in the theory of the
usual spherical model. \cite{berkac52,joy72pt}
The meanings of both equations are, however, different.
Whereas in the standard spherical model a similar equation account for the pretty
unphysical global spin constraint, Eq.  (\ref{sphereq0}) here is, in fact, the
normalization condition $\langle {\bf s}_{\bf r}^2 \rangle = 1$ for the spin
vectors on each of the lattice sites ${\bf r}$ [see Eq. (\ref{statavr})].
Indeed, calculating the spin autocorrelation function in the form symmetrized
over sublattices with the help of Eqs.  (\ref{defcfs}),  (\ref{orthcomp}), and
(\ref{sigcfOZ}), one obtains
%
\begin{eqnarray}\label{constr}
&&
\langle {\bf s}_{\bf r}^2 \rangle =  v_0\!\!\!\int\!\!\!\frac{d{\bf q}}{(2\pi)^d}
\frac 13 \sum_l s^{ll}({\bf q})
\nonumber\\
&&
\qquad
{} =  v_0\!\!\!\int\!\!\!\frac{d{\bf q}}{(2\pi)^d} 
\frac 13 \sum_n \sigma^n ({\bf q}) = 
D \tilde\Lambda_{\alpha\alpha} \bar P .
\end{eqnarray}
That is, the spin-normalization condition is automatically satisfied in our
theory by virtue of Eq.  (\ref{sphereq0}). 
After  $\tilde\Lambda_{\alpha\alpha}$ has been found from this equation, the spin CF's are readily
given by Eqs. (\ref{sigcfOZ}) and (\ref{defcfs}).

To avoid possible confusion, we should mention that in the paper of Reimers,
Ref. \onlinecite{rei92ga}, where Eq. (\ref{sigcfOZ}) with the bare cumulant  
$\Lambda_{\alpha\alpha} = 1/D$ has been obtained, the theoretical approach has
been called the ``Gaussian approximation (GA)''.
This term taken from the conventional theory of phase transitions based on the
Landau free-energy functional implies that the Gaussian fluctuations of the {\em
order parameter} are considered.
In the microscopic language, this merely means calculating correlation
functions of fluctuating spins after applying the MFA.
Such an approach is known to be inconsistent, since correlations are taken
into account after they had been neglected.
As a result, for the {\em kagom\'e} lattice one obtains a phase transition at the temperature 
$T_c = T_c^{\rm MFA} = 2J/D$ but immediately finds that the approach breaks
down below $T_c$ because of the infinitely strong fluctuations.
In contrast to this MFA-based approach, the self-consistent Gaussian
approximation used here allows, additionally, to the Gaussian fluctuations of
the {\em molecular field}, which renormalize $\Lambda_{\alpha\alpha}$ and lead
to the absence of a phase transition for this class of systems.
The SCGA is, in a sense, a ``double-Gaussian'' approximation: The diagram
series in Fig. \ref{kag_scga}$a$ allows for the Gaussian fluctuations of the
order parameter, whereas that in Fig. \ref{kag_scga}$b$ describes Gaussian fluctuations of the 
molecular field.

To close this section, let us work out the expressions for the energy and the
susceptibility of the {\em kagom\'e} antiferromagnet.
For the energy of the whole system, using Eqs. (\ref{dhamdiag}) and (\ref{defcfsig}), as
well as the equivalence of all spin components, one obtains
%
\begin{equation}\label{Utot}
U_{\rm tot} = \langle {\cal H } \rangle  = -\frac{ N }{ 2 } \sum_n
 v_0\!\!\!\int\!\!\!\frac{d{\bf q}}{(2\pi)^d}
 \tilde V_{\bf q}^n  \sigma^n ({\bf q}).
\end{equation}
To obtain the energy per spin $U$, one should divide this expression by $3N$.
With the use of Eq.  (\ref{sigcfOZ}), the latter can be expressed through the
lattice Green's function $\bar P$ of Eq. (\ref{LbarP});  then with the help of
Eq. (\ref{sphereq0}) it can be put into the final form
%
\begin{equation}\label{U}
U = \frac T2 \left( D - \frac{ 1 }{  \tilde\Lambda_{\alpha\alpha} } \right).
\end{equation}
The susceptibility per spin symmetrized over sublattices can be expressed through the
spin CF's as
%
\begin{equation}\label{chidef}
\chi_{\bf q} = \frac{ 1 }{ 3DT } \sum_{ll'} s^{ll'}({\bf q}) .
\end{equation}
With the use of Eq. (\ref{defcfs}) this can be rewritten in the form
%
\begin{equation}\label{chisig}
\chi_{\bf q} = \frac{ 1 }{ 3DT } \sum_n W_n^2({\bf q}) \sigma^n({\bf q}),
\qquad W_n({\bf q}) \equiv \sum_l U_{l n}({\bf q}) ,
\end{equation}
where the diagonalized CF's are given by  Eq. (\ref{sigcfOZ}).
From Eq. (\ref{Uqsm}) it follows that in the limit ${\bf q} \to 0$ one has
$W_1=W_2=0$ and $W_3=\sqrt{3}$.
Thus the homogeneous susceptibility $\chi \equiv \chi_0$ simplifies to
%
\begin{equation}\label{chihomo}
\chi = \frac{ 1 }{ DT } \sigma^3 (0).
\end{equation}
As we shall see in the next section, disappearance of the terms with $n=1$ and
2
from this formula ensures the nondivergence of the homogeneous susceptibility
of the {\em kagom\'e} antiferromagnet in the limit $T\to 0$.
The situation for ${\bf q}\ne 0$ is much more intricate and it will be
considered below in relation to the neutron-scattering cross section.

\section{Thermodynamics of the kagom\'e antiferromagnet}
\label{secThermod}

To put the results obtained above into the form explicitly well behaved in the
large-$D$ limit and allowing a direct comparison with the results obtained by
other methods for systems with finite values of $D$, it is convenient to use
the mean-field transition temperature  $T_c^{\rm MFA} = 2J/D$ as the energy
scale.
With this choice, one can introduce the reduced temperature $\theta$ and the
so-called gap parameter $G$ according to
%
\begin{equation}\label{defthetaG}
\theta \equiv \frac{ T }{ T_c^{\rm MFA} }, 
\qquad G \equiv \frac D\theta  \tilde\Lambda_{\alpha\alpha} .
\end{equation}
In these terms, Eq. (\ref{sphereq0}) rewrites as
%
\begin{equation}\label{sphereq}
\theta G \bar P(G) = 1 
\end{equation}
and determines $G$ as function of $\theta$.
Here $ \bar P(G)$ is defined by Eq. (\ref{LbarP}), where
%
\begin{equation}\label{Pn}
P_n =  v_0\!\!\!\int\!\!\!\frac{d{\bf q}}{(2\pi)^d}
 \frac{ 1 }
{  1 -  G \nu_n ({\bf q}) } , \qquad P_1 = \frac{ 1 }{ 1 - G } ,
\end{equation}
and the reduced eigenvalues $ \nu_n ({\bf q})$ are given by Eq. (\ref{nudef}).
The $\sigma$ CF's of Eq. (\ref{sigcfOZ}), which are proportional to the integrands of
$P_n$, can be rewritten in the form
%
\begin{equation}\label{sigcf}
\sigma^n ({\bf q}) = \frac{ \theta G }{  1 -  G \nu_n ({\bf q}) } .
\end{equation}
Further, it is convenient to consider the reduced energy per spin defined by
%
\begin{equation}\label{defUtil}
\tilde U \equiv U/|U_0|, \qquad U_0 = - J,
\end{equation}
where $U_0$ is the energy per spin at zero temperature.
With the help of Eq. (\ref{U}) $\tilde U$ can be written as
%
\begin{equation}\label{Util}
\tilde U = \theta - 1/G .
\end{equation}
The homogeneous susceptibility $\chi$ of Eq. (\ref{chihomo}) can be rewritten
with the help of Eq. (\ref{nu23qsm})  in the reduced form
%
\begin{equation}\label{chitilhom}
\tilde \chi \equiv 2J \chi = \frac{ G } { 1 + 2G } .
\end{equation}

\begin{figure}[t]
\unitlength1cm
\begin{picture}(11,7)
\centerline{\epsfig{file=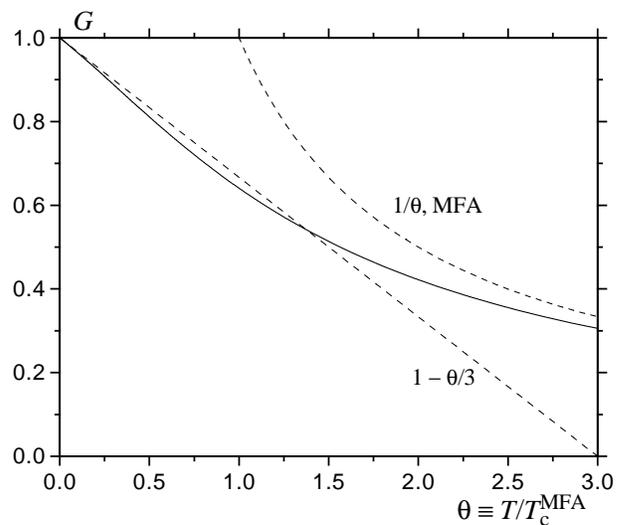,angle=-90,width=12cm}}
\end{picture}
\caption{ \label{kag_GvsT} 
Temperature dependence of the gap parameter $G$ for the {\em kagom\'e} antiferromagnet. 
}
\end{figure}

The sense of calling $G$ the ``gap parameter'' is clear from Eq. (\ref{sigcf}).
If $G=1$, then the gap in correlation functions closes:
$\sigma^1$ turns to infinity, and $\sigma^2$
diverges at $q\to 0$.
For nonordering models, it happens only in the limit $\theta\to 0$,  however. 
The solution of Eq. (\ref{sphereq}) satisfies $G\leq 1$
and goes to zero at high  temperatures. 
If $\theta \ll 1$, the function $\bar P$ is dominated by $P_1 = 1/(1-G)$,
whereas $P_3$ remains of order unity and $P_2$ diverges only logarithmically,
as in usual two-dimensional systems: $P_2 \cong (\sqrt{3}/\pi)\ln[c/(1-G)]$,
$c\sim 1$.
The ensuing asymptotic form of the gap parameter at low temperatures reads
%
\begin{equation}\label{GTlo}
G \cong 1 - \frac \theta 3 - \left( \frac \theta 3 \right)^2 
\frac{ \sqrt{3} }{ \pi } \ln \frac{ 3c }{ \theta } , \qquad \theta \ll 1 .
\end{equation}
At high temperatures, Eq. (\ref{sphereq}) requires small values of $G$.
Here, the limiting form of $\bar P$ can be shown to be $\bar P \cong 1 +
G^2$.
The corresponding asymptote of $G$ has the form
%
\begin{equation}\label{GThi}
G \cong \frac 1 \theta  \left( 1 - \frac{ 1 }{\theta^2 } \right) , \qquad \theta \gg 1 .
\end{equation}
The numerically calculated temperature dependence of $G$ is shown in
Fig. \ref{kag_GvsT}.
Note that in the MFA one has $G=1/\theta$ which attains the value 1 at $\theta=1$.

\begin{figure}[t]
\unitlength1cm
\begin{picture}(11,7)
\centerline{\epsfig{file=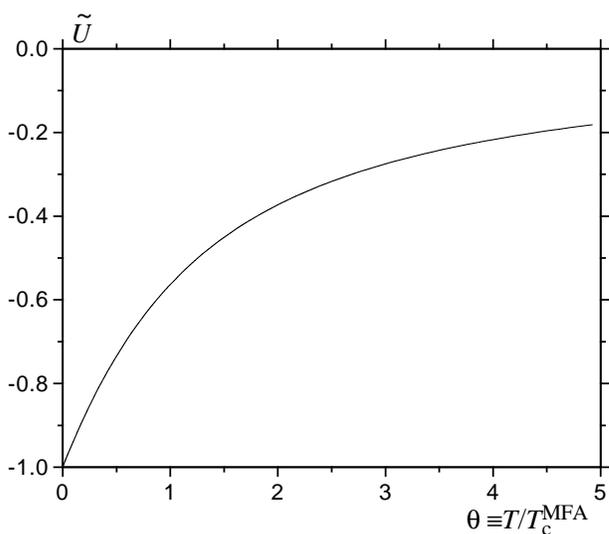,angle=-90,width=12cm}}
\end{picture}
\caption{ \label{kag_UvsT} 
Temperature dependence of the reduced energy of the {\em kagom\'e} antiferromagnet. 
}
\end{figure}
\begin{figure}[t]
\unitlength1cm
\begin{picture}(11,7)
\centerline{\epsfig{file=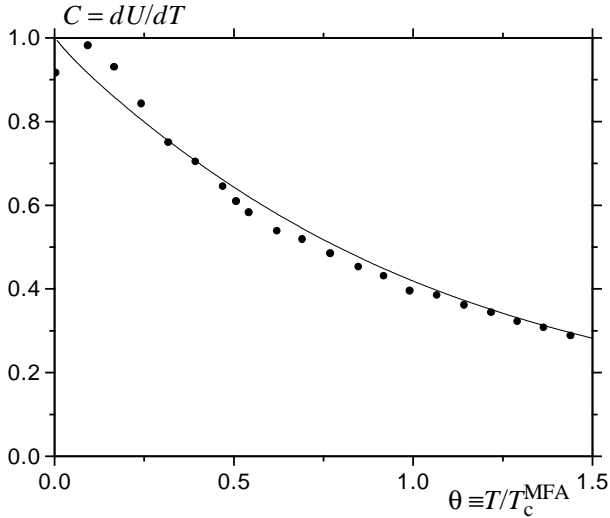,angle=-90,width=12cm}}
\end{picture}
\caption{ \label{kag_CvsT} 
Temperature dependence of the  heat capacity of the {\em kagom\'e}
antiferromagnet. 
The MC results of Ref. \protect\onlinecite{chaholshe92} for the Heisenberg model
($D=3$) are represented by circles.
}
\end{figure}
\begin{figure}[t]
\unitlength1cm
\begin{picture}(11,7)
\centerline{\epsfig{file=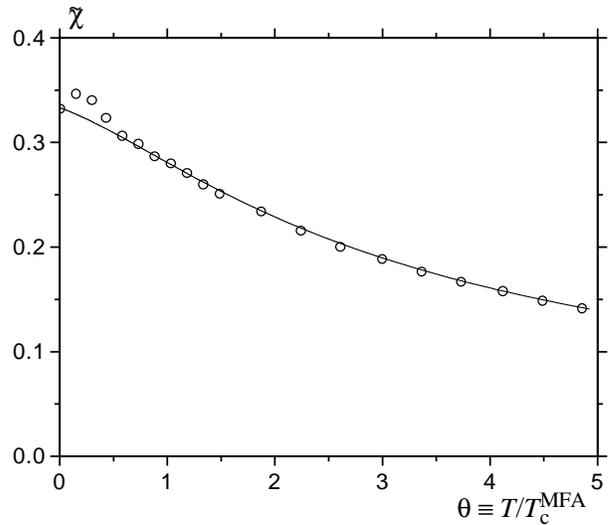,angle=-90,width=12cm}}
\end{picture}
\caption{ \label{kag_XvsT} 
Temperature dependence of the reduced uniform susceptibility of the {\em kagom\'e}
antiferromagnet. 
The MC results of Ref. \protect\onlinecite{reiber93} for the Heisenberg model
($D=3$) are represented by circles.
}
\end{figure}

The temperature dependence of the reduced energy of Eq. (\ref{Util}) is shown
in Fig. \ref{kag_UvsT}.
Its asymptotic forms following from Eqs. (\ref{GTlo}) and (\ref{GThi}) are
given by
%
\begin{equation}\label{UtilAs}
\tilde U \cong \left\{
\begin{array}{ll}
 -1/\theta, 	& \theta \gg 1 \\
-1 + (2/3)\theta,	& \theta \ll 1 .
\end{array}
\right.	 
\end{equation}
This implies the reduced heat capacity $\tilde C = d\tilde U/d\theta$ is
equal to 2/3 at low temperatures, in contrast to $\tilde C = 1$ for usual two-dimensional  lattices in the same approximation. 
The latter result is solely due to the term linear in $\theta$ in Eq. (\ref{Util}), whereas $G$ only 
exponentially deviates from 1 at low temperatures.  
For the {\em kagom\'e} lattice, there is a linear in $\theta$ contribution to the gap parameter $G$ of  Eq. (\ref{GTlo}), which
leads to $\tilde C = 2/3$. 
This reflects the fact that one of three modes in the {\em kagom\'e} lattice [see
Eq. (\ref{dhamdiag})] is dispersionless, and hence 1/3 of all spin degrees
of freedom in the system are local and free, making no contribution to the
heat capacity.

The reduced variables introduced at the beginning of this section are very convenient for the comparison of the 
results for $D=\infty$ with those for finite values of $D$, which are obtained by other methods.
The expected discrepancies are of order $1/D$ which is not too much for $D=3$.
(Note that the $D=\infty$ approximation can be improved by the $1/D$ expansion. \cite{gar94jsp,gar96jsp})
To compare with the MC simulation data of Ref. \onlinecite{chaholshe92} for the heat capacity of the Heisenberg
model we will use, instead of  $\tilde C$, the true heat capacity 
$C = dU/dT = (D/2)\tilde C$ [see Eqs. (\ref{defthetaG}) and (\ref{defUtil})],  which in our approach tends to
$D/3\Rightarrow 1$ 
at low temperatures.
The fairly good agreement on the  high-temperature side of Fig. \ref{kag_CvsT} is not surprising,
since a nontrivial dependence on $D$ appears only at order $1/T^3$ for
the nn correlation function and hence for the energy, and thus at order
$1/T^4$ for the heat capacity [see the combination $n+2 \equiv D+2$ in
Eq. (3.10a) of Ref. \onlinecite{harkalber92}]. 
The reasonable agreement with the MC results at low temperatures is better than
expected and can be interpreted as a compensation of errors.
Indeed, for finite values of $D$ one should take into account the $1/D$
corrections to the present $D=\infty$ results.
For conventional magnets, this leads in the first order in $1/D$ to the replacement of 
$C = D/2$ by $C = (D-1)/2$ in the limit $T\to 0$. \cite{gar94jsp,gar96jsp}
This result is exact and physically transparent as following from the constraint $|{\bf s}_{\bf r}|=1$  in counting 
of the spin degrees of freedom; it does not change further at higher orders of the $1/D$ expansion. 
For the {\em kagom\'e} lattice, the same counting argument suggests to replace $D$ by $D-1$,
which would yield  $C = (D-1)/3 \Rightarrow 2/3$ for $T\to 0$.
On the other hand, inclusion of the $1/D$ corrections reduces the degeneracy of
the ground state, and the heat capacity should increase again.
This degeneracy reduction  manifests itself by the appearance of the ${\bf q}$ dependence of
the correlation function $\sigma_{\alpha\alpha}^1$ of Eq. (\ref{sigcf}).
On the high-temperature side, the degeneracy of the largest eigenvalue of the
susceptibility matrix is removed at order $1/T^8$  [see 
Eqs. (3.29) and (3.31) of Ref. \onlinecite{harkalber92}; the effect vanishes, however,
for $D\to \infty$].
At low temperatures, the resulting heat capacity becomes 11/12
(Ref. \onlinecite{chaholshe92}), which is not far away from our result $C\to 1$.

The reduces uniform susceptibility $\tilde\chi$ calculated from Eq. (\ref{chitilhom}) is
shown in Fig. \ref{kag_XvsT}.
Again, our results are in a fairly good agreement with the MC data of
Ref. \onlinecite{reiber93}, which are, in turn, in accord with the
high-temperature series expansion (HTSE)
results of Ref. \onlinecite{harkalber92} (not shown).
Here, in contrast to the heat capacity, our result $\tilde\chi=1/3$, or $\chi = 1/(6J)$,
at $T=0$ is {\em exact}.
This follows from the fact that the zero-temperature susceptibilities of the classical {\em kagom\'e} antiferromagnet
have the same value $\chi = 1/(6J)$ for all directions of the field with
respect to the three spins on a triangle being mutually oriented at 120$^\circ$.
On the contrary, for conventional low-dimensional antiferromagnets, which show  two-sublattice short-range 
correlations, there are $D-1$ susceptibilities transverse to the local orientation of spins,
$\chi_\perp = 1/(2J_0)$, and one longitudinal susceptibility $\chi_\|$ which vanishes in the zero-temperature limit.
After the averaging over all orientations of spins one obtains the exact result  $\chi = (1-1/D)/(2J_0)$
at $T=0$, which differs significantly from that for the $D=\infty$ model.
Taking into account the first-order $1/D$ correction leads to a rather accurate result for $\chi(T)$ in the whole 
temperature range, \cite{gar94jsp} which has a well-known flat maximum at $T \lsim J$.    
Returning to the {\em kagom\'e} antiferromagnet, one can state that the $1/D$ corrections to  
the susceptibility are smaller than in the conventional case.
The small maximum of $\tilde\chi$ in the data of Ref. \onlinecite{reiber93} is
probably a $1/D$ effect arising due to the increase of the longitudinal
susceptibility of spins with temperature at low temperatures, similarly to
that in conventional low-dimensional antiferromagnets (see
Ref. \onlinecite{gar94jsp} for details).

\section{Real-space correlation functions}
\label{secCF's}

The long-wavelength, low-temperature behavior of the $\sigma$ correlation
functions of Eq. (\ref{sigcf})  is given, according to Eqs. (\ref{nudef}),
(\ref{nu23qsm}), and (\ref{GTlo}), by 
%
\begin{equation}\label{sigcfTlo}
\sigma^1 \cong 3, \qquad \sigma^2 \cong \frac{ 3\kappa^2 }{ \kappa^2 + q^2 },
\qquad \sigma^3 \cong \frac \theta 3 ,
\end{equation}
where the quantity $\kappa^2=2\theta/3$ in $\sigma^2$ defines the correlation length
%
\begin{equation}\label{xidef}
\xi_c = \frac 1\kappa = \left(\frac{ 3 }{ 2\theta } \right)^{1/2} .
\end{equation}
The appearance of this length parameter implies that the real-space spin CF's
defined, according to Eqs. (\ref{defcfs}) and (\ref{fourier}), by
%
\begin{equation}\label{cfR}
s_{ij}^{ll'} =  v_0\!\!\!\int\!\!\!\frac{d{\bf q}}{(2\pi)^d}
 e^{ i \bbox{\rm q \cdot}( {\bf r}_i^l - {\bf r}_j^{l'}) }
U_{l n}({\bf q}) U_{l' n}({\bf q})
\sigma^n({\bf q})
\end{equation}
decay exponentially at large distances at nonzero temperatures.
In contrast to conventional lattices, divergence of $\xi_c$ at $\theta\to 0$
does not lead here to an extended short-range order, i.e., to strong
correlation at distances $r \lsim \xi_c$.
The zero-temperature CF's are {\em purely geometrical} quantities which are dominated by $\sigma^1$ and have the form
%
\begin{eqnarray}\label{cfRT0}
&&
s_{ij}^{ll'} =  3 v_0\!\!\!\int\!\!\!\frac{d{\bf q}}{(2\pi)^d}
 e^{ i \bbox{\rm q \cdot}( {\bf r}_i^l - {\bf r}_j^{l'}) }
U_{l 1}({\bf q}) U_{l' 1}({\bf q})
\nonumber\\
&&\qquad
{} =  3 v_0\!\!\!\int\!\!\!\frac{d{\bf q}}{(2\pi)^d} \cos[ \bbox{\rm q \cdot}( {\bf r}_i^l - {\bf r}_j^{l'}) ]
\nonumber\\
&&
{} \times \frac{ \sin( {\bf u}_l \bbox{\rm \cdot q}) \sin( {\bf u}_{l'} \bbox{\rm \cdot q}) }
{ \sin^2( {\bf u} \bbox{\rm \cdot q}) + \sin^2( {\bf v}\bbox{\rm \cdot q}) + \sin^2( {\bf w}\bbox{\rm \cdot q}) },
\end{eqnarray}
where, according to Eq. (\ref{eigenvec1}), 
%
\begin{equation}\label{deful}
{\bf u}_1 \equiv {\bf w}, 
\qquad  {\bf u}_2 \equiv {\bf v},
\qquad  {\bf u}_3 \equiv {\bf u}.
\end{equation}
At large distances ${\bf r}_{ij}\equiv {\bf r}_i^l - {\bf r}_j^{l'}$, the small values of $q$ are important in
Eq. (\ref{cfRT0}).
Thus one can expand the sines to the lowest order and use 
$({\bf u}\bbox{\rm \cdot q})^2 + ({\bf v}\bbox{\rm \cdot q})^2 + ({\bf w}\bbox{\rm \cdot q})^2 = \threehalf q^2$.
After that integration can be done analytically and yields the asymptotic result
%
\begin{equation}\label{CFDDI}
s_{ij}^{ll'} \cong \frac{ 2\sqrt{3} }{ \pi } \; \frac{ 
({\bf u}_l\bbox{\rm \cdot}{\bf u}_{l'}) r_{ij}^2 - 2 ({\bf u}_l\bbox{\rm \cdot}{\bf r}_{ij}) ({\bf u}_{l'}\bbox{\rm
\cdot}{\bf r}_{ij}) }{ r_{ij}^4 } 
\end{equation}
for $r_{ij} \gg 1$.
That is, at zero temperature spin CF's decrease at the scale of the lattice spacing and
decay according to a power law $1/r^2$ at large distances.
The form of Eq. (\ref{CFDDI}) is that of the dipole-dipole interaction in a
two-dimensional world.
Here the elementary translation vectors
${\bf u}_l$ associated with each of three sublattices [see Eqs. (\ref{deful}), (\ref{defuv}), and (\ref{defqsq3})]  
play the role of dipole moments.

At nonzero temperatures, an additional exponential decay of the correlation functions appears, which is governed by the
correlation length $\xi_c$ of Eq. (\ref{xidef}). 
For $\theta\ll 1$, the third-eigenvalue term, $n=3$, in Eq. (\ref{cfR}) 
can still be neglected, and one can use the first and second columns of the
long-wavelength form of the diagonalizing matrix $U_{l n}({\bf q})$, 
Eq. (\ref{Uqsm}).
The resulting CF $s^{ll'}({\bf q})$ of Eq. (\ref{defcfs}), which enters
Eq. (\ref{cfR}), has the form
%
\begin{equation}\label{cfkap}
s^{ll'}({\bf q}) \cong 
\frac{
\kappa^2 (-1 + 3 \delta_{ll'})
+ 2({\bf u}_l \bbox{\rm \cdot q}) ({\bf u}_{l'} \bbox{\rm \cdot q})  
}{ \kappa^2+q^2 }.
\end{equation}
Whereas the $\kappa^2$ term in the numerator yields only small contributions
$\propto\theta$ in the real-space CF's, that in the denominator results in the
additional exponentially decaying factor 
%
\begin{equation}\label{FactKap}
\renewcommand{\arraystretch}{1.5}
 \kappa r_{ij} K_1(\kappa r_{ij}) 
\cong  \left\{
\begin{array}{ll}
\displaystyle
 1, 	& \kappa n \ll 1 \\
\displaystyle
\sqrt{ \pi\kappa r_{ij}/ 2 }\; e^{-\kappa r_{ij}},	& \kappa n \gg 1,
\end{array}
\right.	 
\end{equation}
in Eq. (\ref{CFDDI}).

To study real-space correlation functions at distances of the order of the
lattice spacing and to list the particular cases of the general formula
(\ref{CFDDI}), it is convenient to enumerate CF's by the numbers $n_u$ and $n_v$ defined by 
Eq. (\ref{transl}), as is shown in Fig. \ref{kag_str}.
Thus $s_{n_u,n_v}^{ll'}$ is the correlation function of the $l$ sublattice
spin of the ``central'' triangle $(0,0)$ with the $l'$ sublattice spin of the
triangle translated by $(n_u,n_v)$. (Note that $s_{n_u,n_v}^{l'l} \ne
s_{n_u,n_v}^{ll'}$, in general.)
There is a number of several useful
relations between correlation functions.
First, the sum of the CF's
$s_{n,0}^{ll}$ over $l$ at $T=0$ is zero by virtue of Eqs. (\ref{cfRT0}) and (\ref{orthcomp}):
%
\begin{equation}\label{cfsymm}
s_{n,0}^{11} + s_{n,0}^{22} + s_{n,0}^{33} =0, \qquad T = 0 .
\end{equation}
Taking into account the symmetry of the lattice, one can put this relation into
the form of the ``star rule''
%
\begin{equation}\label{starrule}
s_{n,0}^{ll} + s_{0,n}^{ll} + s_{n,n}^{ll} =0, \qquad l = 1,2,3
\end{equation}
for the sum of the correlation functions along the  directions ${\bf u}$, ${\bf v}$, and ${\bf w}$
[see  Eqs. (\ref{defuv}) and (\ref{defqsq3})].
The star rule does not hold at nonzero temperatures, which can be easily seen
from the HTSE for the spin CF's  starting from $1/T^{2n}$ for $s_{n,0}^{11}$ and
$s_{n,0}^{22}$ and from  $1/T^{2n+1}$ for $s_{n,0}^{33}$.
More detailed analysis shows that in the low-temperature region the sum in Eqs.  (\ref{cfsymm}) and (\ref{starrule})
is smaller than $s_{n,0}^{33}$ by a factor of order
$(\kappa n)^2 \ln[1/(\kappa n)]$ at the distances $\kappa n \ll 1$.
Thus the star rule can be used with a good accuracy in the
whole range $\theta \ll 1$.
An additional relation can be found from the condition that at zero
temperature the sum of spins in each of the triangles is zero.
Thus one obtains, e.g., the ``triangle rule''
%
\begin{equation}\label{triagrule}
s_{n_u,n_v}^{l1} + s_{n_u,n_v}^{l2} + s_{n_u,n_v}^{l3} = 0, \qquad T = 0
\end{equation}
for all $l$, $n_u$, and $n_u$, as well as similar relations.

The most nontrivial of the relations between spin CF's is the ``hexagon rule''
%
\begin{equation}\label{hexrule}
s_{\rm hex} \equiv 
\sum_{{\bf r}' \in {\rm hex}} (-1)^\zeta s_{{\bf rr}'}  = \sigma^1 \delta_{{\bf r} \in {\rm hex}}
\end{equation}
for the correlators between  a
site ${\bf r}$ and all the sites ${\bf r}'$ belonging to hexagons, 
which are  taken with alternating signs.
If the site  ${\bf r}$ itself belongs to the hexagon, the right-hand side of Eq.
(\ref{hexrule}) is nonzero and the autocorrelation function  $s_{{\bf rr}}$ in
the sum is taken with the positive sign.
As follows from  Eq. (\ref{sigcf}) and the temperature dependence of the gap
parameter $G$, the quantity $s_{\rm hex}$ changes in this case from 1 at high
temperatures to 3 at low temperatures.
This very deep relation has been derived in Ref. \onlinecite{harkalber92} from
the condition that the largest eigenvalue of the correlation matrix 
$\sigma^1$ is independent of ${\bf q}$.
For models with finite $D$ this condition and hence the hexagon rule
(\ref{hexrule}) is violated only at order $1/T^8$ of the HTSE. \cite{harkalber92}
For our $D=\infty$ model, $\sigma^1$ given by Eq. (\ref{sigcfOZ}) remains dispersionless
at all temperatures, and the hexagon rule is always satisfied.

At long distances, the zero-temperature sublattice-diagonal CF's in the horizontal direction,
which follow from Eq. (\ref{CFDDI}), have the form
%
\begin{equation}\label{cf1122}
s_{n,0}^{11} = s_{n,0}^{22} \cong  \frac{ \sqrt{3} }{ \pi r^2},
\qquad s_{n,0}^{33} \cong  -\frac{ 2\sqrt{3} }{ \pi r^2}
\end{equation}
with $r=2n$.
One can see that relation (\ref{cfsymm}) is satisfied. 
For the spin correlators between the first and second
sublattices along the horizontal line one obtains
%
\begin{equation}\label{cf1221}
\left\{ s_{n,0}^{12}  \atop  s_{n,0}^{21} \right\} 
\cong  -\frac{ 2\sqrt{3} }{ \pi r^2 },
\qquad r =  \left\{ 2n+1 \atop  2n-1 \right\} .
\end{equation}
Comparing Eqs. (\ref{cf1122}) and (\ref{cf1221}), one concludes that
correlations in the $D=\infty$ model have nothing in common with the $\sqrt{3}
\times \sqrt{3}$ structure which is selected by thermal fluctuations in the
Heisenberg model. \cite{chaholshe92,husrut92,reiber93} 
Apart from the fast decay, the sign of the correlation function changes with
each step along the line connecting the sites, while the coefficient alternates by the factor 2. 
Such a behavior of the sign and coefficient cannot be described by any
ordering wave vector.
Correlators involving the third sublattice have the form
%
\begin{equation}\label{cf123}
s_{n,0}^{13} \cong s_{n,0}^{31} \cong s_{n,0}^{23} \cong s_{n,0}^{32} \cong 
 \frac{ \sqrt{3} }{ \pi r^2} .
\end{equation}
One can see that the above expressions satisfy the triangle rule, Eq. (\ref{triagrule}). 
In addition, the CF's along the three
lines going through the apexes of the David stars in Fig. \ref{kag_str}
read
%
\begin{equation}\label{cfDavid}
s_{n,-n}^{11} = s_{2n,n}^{22} = s_{n,2n}^{33} \cong 
 \frac{ 2\sqrt{3} }{ \pi r^2 } .
\end{equation}

\begin{figure}[t]
\unitlength1cm
\begin{picture}(11,7)
\centerline{\epsfig{file=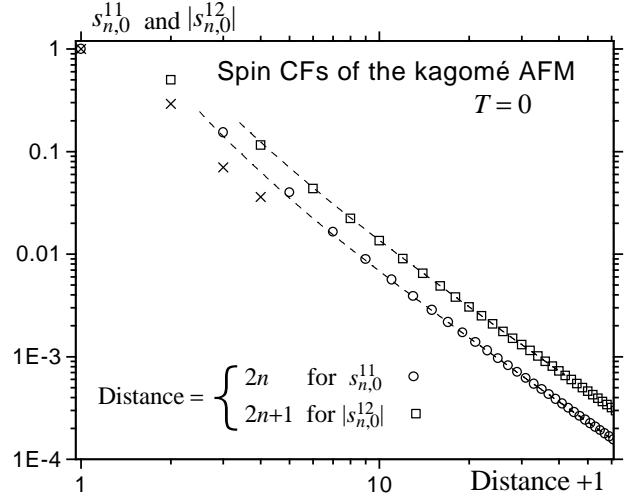,angle=-90,width=12cm}}
\end{picture}
\caption{ \label{kag_cf} 
Real-space correlation functions $s_{n,0}^{11}$ and  $|s_{n,0}^{12}|$  at $T=0$
calculated from Eq. (\protect\ref{cfRT0}). 
The distance unit is the interatomic spacing.
The asymptotes given by Eqs. (\protect\ref{cf1122}) and (\protect\ref{cf1221})
are shown by the dashed lines.
Crosses are the results of Ref. \protect\onlinecite{leuels93} for the quantum system
with $S=1/2$ multiplied by 4.
}
\end{figure}

To calculate the short-range correlation functions, one should use in 
 Eqs. (\ref{cfR}) or (\ref{cfRT0}) the full
form of the diagonalizing matrix $U_{ln}({\bf q})$ [see Eq. (\ref{eigenvec})]
instead of its long-wavelength form (\ref{Uqsm}) and integrate over the whole
Brillouin zone.
This seems to be impossible to do analytically, but at $T=0$ one can express numerous
CF's through some ``fundamental'' one with the help of the relations discussed
above. 
So, in addition to the trivial results $s_{0,0}^{ll}=1$,  
$s_{0,0}^{12}=s_{1,0}^{21}=-1/2$, etc., one obtains numerically
%
\begin{equation}\label{cffund}
s_{1,0}^{11} = s_{1,0}^{22} = a = 0.1540 .
\end{equation}
After that using the star and triangle rules leads to the results
%
\begin{eqnarray}\label{cfsho1}
&&
s_{1,0}^{33} = -2a = -0.308 		
\nonumber\\
&&
s_{1,1}^{13} = -a + 1/2 = 0.346
\nonumber\\
&&
s_{1,1}^{12} = s_{1,0}^{13} = 3a - 1/2 = -0.038
\nonumber\\
&&
s_{1,0}^{12} = -4a +1/2 = 0.116
\nonumber\\
&&
s_{1,-1}^{11} = -6a +1 = 0.076
\nonumber\\
&&
s_{1,-1}^{12}  = 3a - 1/2 = -0.038 .
\end{eqnarray}
Now from the hexagon rule for the hexagon marked by the star in Fig. \ref{kag_str},
%
\begin{equation}\label{hexrule2}
s_{\rm hex} = s_{1,0}^{11} - s_{1,1}^{13} + s_{1,1}^{12} -  s_{2,1}^{11} +
s_{2,1}^{13} - s_{1,0}^{12} = 0,
\end{equation}
and from other relations one obtains the CF's on the remote side of this hexagon:
%
\begin{eqnarray}\label{cfsho2}
&&
s_{2,1}^{11} =  3a - 1/2 = -0.038		
\nonumber\\
&&
s_{2,1}^{13} =  -6a +1 = 0.076 .
\end{eqnarray}
After that the star and triangle rules yield
%
\begin{eqnarray}\label{cfsho3}
&&
s_{2,0}^{11} =  10a - 3/2 = 0.040		
\nonumber\\
&&
s_{2,0}^{12} =  -36a + 11/2 = -0.044
\nonumber\\
&&
s_{2,0}^{13} =  26a - 4 = 0.004
\nonumber\\
&&
s_{2,-1}^{11} =  -29a +9/2 = 0.034 .	
\end{eqnarray}
This routine cannot be continued without numerically calculating the next
fundamental CF $s_{3,0}^{11} = 0.0164$.
This would make little sense, however, because at such distances correlation
functions are already well described by their asymptotic forms given above (see Fig. \ref{kag_cf}). 

The results obtained above for the real-space CF's can be compared with
those for the quantum Heisenberg antiferromagnet with $S=1/2$. \cite{leuels93} 
The latter for the CF's of the same type,
multiplied by a factor of four to normalize the autocorrelation function by
one, are also shown in Fig. \ref{kag_cf}.
In contrast to the classical Heisenberg antiferromagnet on the {\em kagom\'e}
lattice selecting the $\sqrt{3}\times\sqrt{3}$ structure, the ground state of
the quantum model is disordered,  \cite{zenels90,leuels93} which brings it, in
a sense, closer to our $D=\infty$ model.
In the quantum model CF's decay faster: Even for the nearest neighbors one has  
 $s_{0,0}^{12} = -0.2922$ instead of $-0.5$ because of the zero-point motion.
For the establishing of the large-distance behavior of the CF's in the quantum
antiferromagnet, the numerical diagonalization of clusters of much larger
sizes (36 sites in Ref. \onlinecite{leuels93} or 27 sites in
Ref. \onlinecite{lecetal97}) is needed, which is a tremendous computational problem.

The main implication of this section is that the spin correlation functions for
the large-$D$ model on the {\em kagom\'e} lattice decay at the distances of the
order of the lattice spacing even at $T\to 0$, in spite of the divergence of
the correlation length $\xi_c$. 
Thus $T=0$ is {\em not a critical point} of
this model, in contrast to the conventional low-dimensional ferro- and antiferromagnets.
Correlations developing in the low-temperature range, which, however, become
only strong between several neighboring spins, characterize the state of
our model as a {\em spin liquid}.

\section{Neutron-scattering cross section}
\label{secNeutron}

The static magnetic neutron-scattering cross section is proportional to the
static Fourier-transformed spin CF:
%
\begin{equation}\label{defcrsec}
\frac{ d\sigma }{ d\Omega } \propto \sum_{{\bf rr}'} 
\langle S_{\bf r}^\perp S_{{\bf r}'}^\perp \rangle e^{i{\bf q}({\bf r} - {\bf r}')},  
\end{equation}
where $S_{\bf r}^\perp$ is the component of the spin perpendicular to the
scattering wave vector ${\bf q}$.
In our model all spin components are equivalent and ${\bf r} = {\bf r}_i^l$ is
defined in Sec. \ref{secStructure}.
Since the
overall coefficient in Eq. (\ref{defcrsec}) contains a magnetic form factor and is
poorly known, one can use the most convenient form of this coefficient and
define
%
\begin{equation}\label{crsec}
\frac{ d\sigma }{ d\Omega } =  \frac 13 \sum_{ll'} s^{ll'}({\bf q})  =
 \frac 13 \sum_n W_n^2({\bf q}) \sigma^n({\bf q}).
\end{equation}
This expression differs from the wave-vector-dependent susceptibility of
Eqs. (\ref{chidef}) and (\ref{chisig})  only by the absence of $DT$ in the denominator,
while the normal-mode CF's $\sigma^n$ are given by Eq. (\ref{sigcf}).
The scattering wave vector ${\bf q}$ is not confined to the Brillouin zone (BZ), in
contrast to ${\bf q}$ appearing in the calculation of the thermodynamic
quantities and real-space CF's.
For usual bipartite lattices, the scattering  cross section with ${\bf q}$ outside the
BZ is the same as with ${\bf q}' ={\bf q}- \bbox{\kappa} $ inside the
BZ, where $\bbox{\kappa}$ is an appropriate reciprocal-lattice
vector.
That is, in this case $d\sigma/d\Omega$ is repeating over the set of extended
Brillouin zones. 
For periodic lattices with more complicated structures, the neutron cross
section is still a periodic function of ${\bf q}$, but the period can be larger than one BZ.

\begin{figure}[t]
\unitlength1cm
\begin{picture}(11,8)
\centerline{\epsfig{file=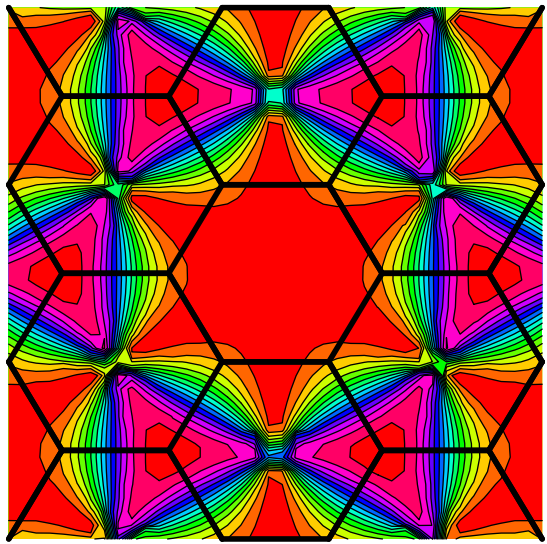,angle=0,width=13.5cm}}
\end{picture}
\caption{ \label{kag_neu} 
Neutron-scattering cross section from the large-$D$ {\em kagom\'e} antiferromagnet at $T=0$
(cf. Fig. \protect\ref{kag_mod}).
}
\end{figure}

For the {\em kagom\'e} lattice in the limit $T\to 0$, Eq. (\ref{crsec}) is dominated
by the term with $\sigma^1 = \theta G/(1-G)$.
Using $G\cong 1- \theta/3$ one obtains $d\sigma/d\Omega = W_1^2({\bf q})$ that is
temperature independent.
As follows from the contour plot in Fig. \ref{kag_neu}, the ``unit cell'' for
the neutron cross section contains four Brillouin zones: one with a very low
scattering intensity, such as the first BZ in the middle, and three BZ's with a
highly inhomogeneous scattering pattern oriented at three different angles.  
The neutron cross section is symmetric with respect to rotations by $60^\circ$
degrees.
It  reaches its maximal value for ${\bf q}=(\pm 8\pi/3, 0)$, etc., 
and vanishes along the directions $q_x=0, \pm\sqrt{3}q_y$, including ${\bf q}=0$.
The scattering pattern in Fig. \ref{kag_neu} strongly resembles that
in the appropriate plane for the classical Heisenberg model on the pyrochlore
lattice, which was obtained by MC simulations in
Ref. \onlinecite{zinharzei97}.
In contrast, the perturbative calculation for the quantum AFM model with $S=1/2$ on the
pyrochlore lattice \cite{canlac98} shows much less revealed triangular shape of the maxima
of the neutron cross section.

\begin{figure}[t]
\unitlength1cm
\begin{picture}(11,7)
\centerline{\epsfig{file=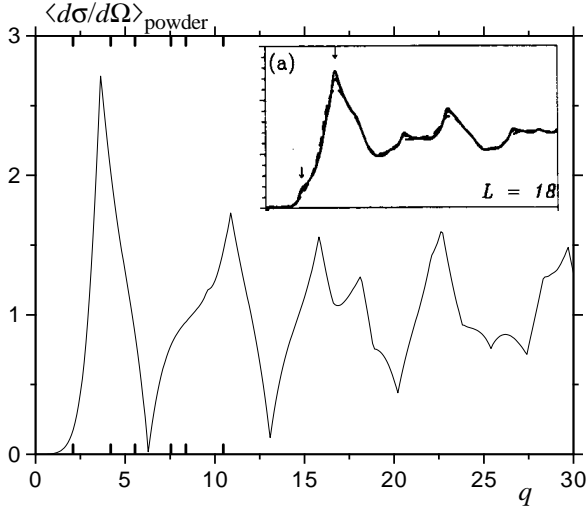,angle=-90,width=12cm}}
\end{picture}
\caption{ \label{kag_powd} 
Powder average of the neutron-scattering cross section from the {\em kagom\'e}
antiferromagnet at $T=0$.
Inset: MC simulations of Ref. \protect\onlinecite{reiber93} for the Heisenberg model. 
}
\end{figure}

Let us now analyze the neutron cross section in more detail.
At $T=0$ using Eqs. (\ref{eigenvec1}) and (\ref{chisig}) one obtains
%
\begin{equation}\label{crsecT0}
\frac{ d\sigma }{ d\Omega } =  
\frac{ 
4 \sin^2(q_x/2) [ \cos(q_x/2) - \cos(\sqrt{3}q_y/2) ]^2
}{
\sin^2(q_x) +\sum_\pm \sin^2(q_x/2\pm\sqrt{3}q_y/2)
},
\end{equation}
where $\sum_\pm$ sums terms with both signs.
Particular cases of this formula are the following.
For $q_y=0$ one has
%
\begin{equation}\label{crsecqy0}
\frac{ d\sigma }{ d\Omega } =  \frac{ 4 \sin^4(q_x/4) }{ 1 + 2  \cos^2(q_x/2) } .
\end{equation}
This is equal to 1/3 at the corner of the first BZ, $q_x=2\pi/3$,  to 2 at
$q_x=\pi$ (the highest slope), to 3 at
$q_x=4\pi/3$ (the maximum),  and to 8/3 at $q_x=2\pi$.
Near the line $q_x=0$,  Eq. (\ref{crsecT0}) simplifies to
%
\begin{equation}\label{crsecqxsm}
\frac{ d\sigma }{ d\Omega } \cong  \frac{ q_x^2 }{ 2 } \tan^2 \frac{ \sqrt{3} q_y}{4 },
\qquad q_y, \left | q_y - \frac{ 2\pi }{ \sqrt{3} } \right |  \gg q_x .
\end{equation}
For small wave vectors one obtains
%
\begin{equation}\label{crsecqsm}
\frac{ d\sigma }{ d\Omega } \cong  \frac{ 1 }{ 96 } 
\frac{ q_x^2 (q_x^2-3q_y^2)^2 }{ q_x^2 + q_y^2 },  \qquad q_x, q_y \ll 1.
\end{equation}
The most interesting form of the neutron cross section is realized in the
vicinity of the centers of the Brillouin zones surrounding the first BZ in
Fig. \ref{kag_neu}.
In particular, near ${\bf q} = (0, 2\pi/\sqrt{3})$ Eq. (\ref{crsecT0}) takes the form
%
\begin{equation}\label{crsecsing}
\frac{ d\sigma }{ d\Omega } \cong  \frac 83
\frac{ q_x^2 }{ q_x^2 + (\delta q_y)^2 },  \qquad q_x, \delta q_y \ll 1,
\end{equation}
where $\delta q_y \equiv q_y - 2\pi/\sqrt{3}$.
This function is nonanalytic at $q_x, \delta q_y =0$, and its limiting value
in this point depends on the way of approaching to.
(Such a function is difficult to plot: What looks like narrow paths in
Fig. \ref{kag_neu} are in fact infinitely thin walls.)

At $T\ne 0$ one should take into account the terms with $n=2,3$ in
Eq. (\ref{crsec}).
This is especially important in the region where the zero-temperature neutron cross section 
turns to zero or is singular.
In particular, near ${\bf q}= 0$ the quantity $W_1^2$ is small due to
cancellation of the leading terms and is given by the right-hand side (rhs) of Eq. (\ref{crsecqsm}).
Similar cancellation takes place in $W_2^2$, and the result is given by the
same formula with interchanged $q_x$ and $q_y$.
The leading term thus becomes the noncancelled one associated with the ``optical''
eigenvalue,  $W_3^2 = 3$.
This leads to the uniform susceptibility given by Eqs. (\ref{chihomo}) and  (\ref{chitilhom}),
that tends to a constant in the low-temperature limit.
On the contrary, $\chi_{\bf q}$ with ${\bf q}\ne 0$ behaves as $1/T$ at low
temperatures.

In the vicinity of the singularity point ${\bf q} = (0, 2\pi/\sqrt{3})$, i.e.,
near the center of the Brillouin zone just above the first (central) BZ in
Fig. \ref{kag_neu},
cancellation of the leading terms does not take place.
Here the matrix $\hat U_{\bf q}$ differs from that of Eq. (\ref{Uqsm}) by the
redefinition $n_y \equiv \delta q_y/[q_x^2+(\delta q_y)^2]^{1/2}$ and by the
change of sign of the third row.
Here $W_2^2$ is given by the rhs of  Eq. (\ref{crsecsing}) with interchanged
$q_x$ and $\delta q_y$.
Now with the use of Eq. (\ref{sigcfTlo}) one obtains the final result
%
\begin{equation}\label{crseckap}
\frac{ d\sigma }{ d\Omega } \cong  \frac 83
\frac{ \kappa^2 + q_x^2 }{ \kappa^2 + q_x^2 + (\delta q_y)^2 },  
 \qquad \kappa, q_x, \delta q_y \ll 1.
\end{equation}
It can be seen that this expression is nonanalytic only in the limit
$T\to 0$, where the correlation length $\xi_c$ defined by Eq. (\ref{xidef})
becomes infinite.

The powder average of the neutron cross section $\langle d\sigma/d\Omega \rangle$,
i.e., the average of
Eq. (\ref{crsec}) over the directions of ${\bf q}$, is shown in Fig. \ref{kag_powd}.
Positions of its singularities at $T=0$ can be found from the scattering pattern in
Fig. \ref{kag_neu}.
The first of them are located at 
$q = 2\pi/\sqrt{3}\approx 3.63$ (sharp maximum),
$2\pi \approx 6.28$ (sharp minimum), 
$\sqrt{(3\pi)^2 + (\pi/\sqrt{3})^2} = 2\pi\sqrt{7/3} \approx 9.6$ (small
cuspy shoulder), 
$ 3\times 2\pi/\sqrt{3} \approx 10.88$ (sharp maximum), 
$\sqrt{(4\pi)^2 + (2\pi/\sqrt{3})^2} = 2\pi\sqrt{13/3} \approx 13.08$ (sharp
minimum),
$\sqrt{(5\pi)^2 + (\pi/\sqrt{3})^2} = 2\pi\sqrt{19/3} \approx 15.81$ (sharp
maximum),
$ 5\times 2\pi/\sqrt{3} \approx 18.14$ (sharp maximum), 
$6\pi \approx 18.85$ (cuspy shoulder), 
etc.
With increasing of $q$ the behavior of $\langle d\sigma/d\Omega \rangle$ becomes
more and more irregular, and it very slowly approaches the value 1.
The latter can be understood since for large values of $q$ the average over the
directions of ${\bf q}$ should be equal to that over ${\bf q}$ itself.
The latter is according to Eq. (\ref{defcrsec}) nothing else but the
autocorrelation function, and the result is unity for the normalization of  $d\sigma/d\Omega$
adopted in Eq. (\ref{crsec}).

At nonzero temperatures, the sharp features of $\langle d\sigma/d\Omega\rangle$ smoothen.
Their low-temperature forms can be found with the help of Eq. (\ref{crseckap}) and are given by
%
\begin{equation}\label{max1}
\left\langle \frac{ d\sigma }{ d\Omega }\right\rangle \cong 2.712 + 0.868 \delta q -
 \frac{ 4\sqrt{3} } { \pi } \frac{  (\delta q)^2 }{ \sqrt{ \kappa^2 + (\delta q)^2 } }
\end{equation}
near the first maximum, $\delta q \equiv q- 2\pi/\sqrt{3}$, and
%
\begin{equation}\label{min1}
\left\langle \frac{ d\sigma }{ d\Omega }\right\rangle \cong 0.0183 -0.238 \delta q
+ \frac{ 4 } { \pi } \sqrt{ \kappa^2 + (\delta q)^2 } 
\end{equation}
near the first minimum, $\delta q \equiv q- 2\pi$.
In the high-temperature limit,  Eq. (\ref{defcrsec}) is dominated by the
autocorrelation function, and the neutron cross section defined by
Eq. (\ref{crsec}) is equal to 1 for all ${\bf q}$ (an absolutely diffuse
scattering).

The MC simulation data  of Ref. \onlinecite{reiber93} for the Heisenberg model
at $T=0.002 J$ are shown in the inset to Fig. \ref{kag_powd}.
At such low temperatures the system shows a tendency towards selection of 
the $\sqrt{3}\times\sqrt{3}$ phase, and the corresponding Bragg-reflection
peaks grow.
The latter, according Eq. (\ref{defqsq3}), are situated at the corners of all 
Brillouin zones in the extended BZ scheme shown in Fig. \ref{kag_neu}, and their
positions are shown by additional tics in Fig. \ref{kag_powd}.
These peaks that are superimposed on the underlying $D=\infty$ structure
 can be traced out in the inset.
Note that there are  Bragg-reflection peaks in the vicinity of the $D=\infty$
peaks, and they seem to be mixed together in the simulations of
Ref. \onlinecite{reiber93}.
In contrast, the first two minima in Fig. \ref{kag_powd} can be found in the
inset at nearly the same positions, although in a strongly rounded form.

\section{Discussion}
\label{secDiscussion}

In the main part of this article, we have presented in detail the exact
solution for the $D=\infty$ component classical antiferromagnet on the
{\em kagom\'e} lattice.
The solution does not show ordering at any temperature due to the strong
degeneracy of the ground state, and the thermodynamic functions behave smoothly. 
In contrast to conventional two-dimensional magnets, there is no extended
short-range order at low temperatures, and $T=0$ is not a critical point of
the system. 
Although the correlation length diverges as $\xi_c \propto T^{-1/2}$, the power-law
decay $\langle {\bf s}_0 {\bf s}_{\bf r} \rangle \propto 1/r^2$ of the
spin correlation functions leads to the loss of correlations at the scale of the
interatomic distance.
The magnetic neutron-scattering cross section becomes nonanalytic at $T\to 0$
but does not diverge at any ``ordering'' wave vector.

Although the model with an infinite number of spin components may appear very
unphysical at the first glance, it is in fact the second that should be
applied, after the mean-field approximation, to any classical spin system.
It properly describes the effect of would be Goldstone modes and thus it has
important advantages against the MFA.
Properly scaled 
physical quantities show a smooth dependence on $D$, and in typical cases
the large-$D$ model proves to be a reasonable approximation to the realistic
one with $D=3$.
So, the results for the heat capacity and the uniform susceptibility obtained
above are in a fairly good agreement with the MC simulation results for the
Heisenberg model in the whole temperature range.  
This implies that the $1/D$ corrections to the thermodynamic functions of the
{\em kagom\'e} AFM, which could be studied within the same
theoretical framework, \cite{gar94jsp,gar96jsp} are suppressed by some mechanism.

The $D=\infty$ model and the $1/D$ expansion seem to be inefficient in the
cases when, due to topological effects, the behavior of the system abruptly
changes at small values of $D$.
A well-known example is the Berezinskii-Kosterlitz-Thouless transition which
takes place in two dimensions for $D=2$.
For the {\em kagom\'e} lattice,  thermal fluctuations favor the $\sqrt{3}\times \sqrt{3}$ phase at low 
temperatures for the Heisenberg model, but there is no
such an effect for $D>3$.  \cite{husrut92,moecha98}
On the other hand, the tendency to the selection of
the $\sqrt{3}\times \sqrt{3}$ phase at low temperatures is already seen in the
high-temperature series expansion of Ref. \onlinecite{harkalber92} for any finite value of $D$. 
Exactly how this mechanism becomes inefficient at low temperatures for $D>3$, could be
studied with the help of the $1/D$ expansion.
The latter describes lifting of the degeneracy of the largest eigenvalue of
the correlation matrix in the first order in $1/D$ and is applicable in the
whole range of temperatures.

As follows from the consideration above, the Heisenberg antiferromagnet on the
{\em kagom\'e} lattice is still not the best model to substitute it with the $D=\infty$
model.
For the Heisenberg AFM on the pyrochlore lattice, the large-$D$ approximation
can be expected to work even better since this model is, in a sense, more disordered, and topological effects leading here to the
selection of the coplanar phase arise only for $D=2$. \cite{moecha98}
The formalism for the pyrochlore lattice in zero field is essentially the same
as for the {\em kagom\'e} lattice, and the corresponding results will be presented elsewhere.

\section*{Acknowledgments}

We would like to thank John Berlinsky for the permission to use the data of
Ref. \onlinecite{reiber93} in the inset to Fig. \ref{kag_powd} and 
Konstantin Kladko for a critical reading of the manuscript.
D. G. is grateful to Christopher Henley for a stimulating discussion.


\end{document}